\documentclass[3p,times]{elsarticle}
\pdfoutput=1

\usepackage{graphicx}
\usepackage{amssymb,amsmath}
\usepackage{bbm}
\usepackage{algorithmic}
\usepackage{hyperref}

\newcommand{\Dw}{D_\text{w}}
\newcommand{\Dov}{D_\text{ov}}
\newcommand{\kout}{k}
\newcommand{\kin}{\ell}
\newlength{\figwidth}
\newcommand{\normx}{\lvert x \rvert}
\newcommand{\order}{{\cal O}}
\newcommand{\lopt}{\kin_\text{opt}}

\DeclareMathOperator{\sgn}{sgn}
\DeclareMathOperator{\re}{Re}

\DeclareMathOperator{\myspan}{span}
\DeclareMathOperator{\Gl}{Gl}

\DeclareMathOperator{\diag}{diag}

\begin{document}

\begin{frontmatter}

\title{A nested Krylov subspace method to compute the sign function of large complex matrices\tnoteref{t1}}
\author{Jacques C.R. Bloch}
\author{Simon Heybrock}
\address{Institute for Theoretical Physics, University of
  Regensburg, 93040 Regensburg, Germany}
\tnotetext[t1]{Supported by the DFG collaborative research center SFB/TR-55 ``Hadron Physics from Lattice QCD''.}

\date{December 22, 2009}

\begin{abstract}
We present an acceleration of the well-established Krylov-Ritz methods to compute the sign function of large complex matrices, as needed in lattice QCD simulations involving the overlap Dirac operator at both zero and nonzero baryon density. Krylov-Ritz methods approximate the sign function using a projection on a Krylov subspace. To achieve a high accuracy this subspace must be taken quite large, which makes the method too costly. The new idea is to make a further projection on an even smaller, nested Krylov subspace. If additionally an intermediate preconditioning step is applied, this projection can be performed without affecting the accuracy of the approximation, and a substantial gain in efficiency is achieved for both Hermitian and non-Hermitian matrices. The numerical efficiency of the method is demonstrated on lattice configurations of sizes ranging from $4^4$ to $10^4$, and the new results are compared with those obtained with rational approximation methods.
\end{abstract}

\end{frontmatter}

\section{Introduction}

In quantum chromodynamics (QCD) some physical observables rely on the chiral properties of the theory. To study such observables in a lattice formulation of QCD it is important to discretize the Dirac operator such that it respects the corresponding chiral symmetry. This is most faithfully achieved using the overlap Dirac operator \cite{Narayanan:1993sk,Narayanan:1994gw}.
To study QCD at nonzero baryon density the overlap formulation was recently extended to include a quark chemical potential \cite{Bloch:2006cd,Bloch:2007xi}.
A major ingredient in the overlap operator, which makes its use very challenging, is the computation of the sign function of a complex matrix, which is Hermitian at zero baryon density, but becomes non-Hermitian when a quark chemical potential is introduced.

The search for efficient numerical methods to compute the sign function for the large sparse matrices encountered in this context is an ongoing field of research. 
Typically, Krylov subspace methods are employed to evaluate the operation of a matrix function on an arbitrary vector. 
We distinguish two main variants: the Krylov-Ritz approximation, which 
evaluates the function via a projection on the Krylov subspace, and the rational approximation, where the function is first approximated by a partial fraction expansion, which is then efficiently solved using a multi-shift Krylov subspace inverter. 

In the Hermitian case efficient rational approximation methods for the sign function have been devised \cite{Neuberger:1998my,vandenEshof:2002ms} and are currently being used in large scale lattice simulations. The current method of choice uses the Zolotarev partial fraction expansion \cite{vandenEshof:2002ms,Chiu:2002eh,Kennedy:2004tk}, 
which yields the optimal rational approximation to the sign function over a real interval \cite{zolotarev77}, in conjunction with a multi-shift conjugate gradient inversion. 
For non-Hermitian matrices, which occur in the presence of a quark chemical potential, Krylov subspace approximations to the sign function are relatively new and still under development. Recently, partial fraction expansion methods using the Neuberger expansion \cite{Neuberger:1998my} with non-Hermitian multi-shift inverters were proposed \cite{Bloch:2009in}.

The Krylov-Ritz approximation, which we discuss in this paper, is based on the construction of a Krylov basis and its accompanying Ritz matrix.
Depending on the algorithm used to construct the basis we distinguish between the Lanczos approximation in the Hermitian case \cite{vandenEshof:2002ms}, and the Arnoldi approximation \cite{Bloch:2007aw} or two-sided Lanczos approximation \cite{Bloch:2008gh} in the non-Hermitian case.
The latter clearly yields the more efficient function approximation for non-Hermitian matrices \cite{Bloch:2008gh}.
In the Krylov-Ritz approximation the large complex matrix is projected on the Krylov subspace, and its sign function is approximated by lifting the sign function of its projected image (Ritz matrix) back to the original space. 
The latter sign function is computed to high accuracy using the spectral definition of a matrix function or using a matrix-iterative method. 
When a large Krylov subspace is needed to reach the desired accuracy, the computation of this matrix sign function becomes a bottleneck for the algorithm. 

Herein we will introduce an enhancement of the Krylov-Ritz approximation method which substantially reduces the cost of this internal sign computation and boosts the efficiency of the overall method, such that it competes with, and even surpasses, the rational function approximation in both the Hermitian and non-Hermitian case. 
The dramatic reduction in computation time is achieved by projecting the Ritz matrix on an even smaller, nested Krylov subspace, after performing a suitable preconditioning step first. The desired sign function is then computed via the sign function of the inner Ritz matrix, which yields the same accuracy as the original Krylov-Ritz approximation.

The outline of the paper is as follows. In Sec.~\ref{Sec:definition} we introduce the overlap operator and the matrix sign function. In Sec.~\ref{Sec:Krylov} we show how the matrix function of large matrices is computed using Krylov-Ritz approximation methods. In Sec.~\ref{Sec:nested} we introduce the nested Krylov subspace method, which substantially enhances the efficiency of the Krylov-Ritz approximation to the sign function. We study its convergence properties and present numerical results for various lattice sizes, including a comparison with rational approximation methods. Finally, our conclusions are given in Sec.~\ref{Sec:conclusions}. For completeness we have added some algorithms in Appendix.

\section{Overlap operator and the matrix sign function}
\label{Sec:definition}

Our motivation to develop numerical algorithms to compute the matrix sign function of large, sparse, complex matrices comes from its application in lattice quantum chromodynamics (LQCD).
The overlap formulation of the Dirac operator \cite{Narayanan:1993sk,Narayanan:1994gw}, which ensures that chiral symmetry is preserved in LQCD, 
is given in terms of the matrix sign function \cite{Neuberger:1997fp}, and its definition in the presence of a quark chemical potential $\mu$ \cite{Bloch:2006cd} is given by 
\begin{align}
\Dov(\mu) = \mathbbm{1} + \gamma_5\sgn(\gamma_5
\Dw(\mu)),
\label{Dovmu} 
\end{align}
where $\mathbbm{1}$ denotes the identity matrix, $\gamma_5=\gamma_1\gamma_2\gamma_3\gamma_4$ with
$\gamma_1,\ldots,\gamma_4$ the Dirac gamma matrices in Euclidean
space, $\sgn$ is the matrix sign function, and
\begin{align}
\Dw(\mu)
 = \mathbbm{1} - \kappa \sum_{i=1}^3 ( T_i^+ + T_i^-) 
- \kappa ( {e^\mu} T_4^+ + {e^{-\mu}} T_4^-)
\end{align}
is the Wilson Dirac operator at nonzero chemical potential
\cite{Hasenfratz:1983ba} with $(T^{\pm}_\nu)_{yx} = (\mathbbm{1} \pm
\gamma_\nu) U_{x,\pm\nu} \delta_{y,x\pm\hat\nu}$, $\kappa =
1/(8+2m_\text{w})$, $m_\text{w} \in (-2,0)$ and $U_{x,\pm\nu}\in$
SU(3), where $U_{x,-\nu}={U^\dagger}_{\!\!\!{x-\hat\nu,+\nu}}$.  The exponential factors $e^{\pm\mu}$ implement the quark
chemical potential on the lattice.
For $\mu = 0$ the argument of the sign function is Hermitian, while for $\mu\ne 0$ it is non-Hermitian. To compute the overlap operator we need to define the matrix sign function for a general complex matrix $A$ of dimension $n$. A generic matrix function $f(A)$ can be defined by
\begin{align}
    \label{eq:matfun}
    f(A) &= \frac{1}{2\pi i} \oint_\Gamma f(z)(zI-A)^{-1}dz,
\end{align}
where $\Gamma$ is a collection of contours in $\mathbb{C}$ such that $f$ is analytic inside and on $\Gamma$ and such that $\Gamma$ encloses the spectrum of $A$. If $A$ is diagonalizable, i.e., $A=U\Lambda U^{-1}$, with diagonal eigenvalue matrix $\Lambda = \diag(\lambda_1, \ldots, \lambda_n)$ and $U\in \Gl(n,\mathbb{C})$, then this general definition can be simplified to the well-known spectral form
\begin{align}
    f(A) &= Uf(\Lambda)U^{-1} ,
\label{fAdef}
\end{align}
with
\begin{align}
    f(\Lambda) &= \diag\left(f(\lambda_1),\ldots,f(\lambda_n)\right).
\end{align}
If $A$ cannot be diagonalized, a spectral definition of $f(A)$ can still be derived using the Jordan decomposition \cite{Golub}. For simplicity, but without loss of generality, we assume diagonalizability in the following. 
For Hermitian $A$ the eigenvalues are real and their sign is defined by $\sgn(x)=\pm 1$ for $x\gtrless 0$ with $x \in \mathbb{R}$, such that Eq.~\eqref{fAdef} readily defines the matrix sign function.
For non-Hermitian $A$ the eigenvalues are complex and require a definition of $\sgn(z)$ for $z \in \mathbb{C}$. The sign function needs to satisfy $(\sgn(z))^2=1$ and reproduce the usual $\sgn(x)$ for real $x$. We define
\begin{align}
    \sgn(z) &= \frac{z}{\sqrt{z^2}} = \sgn\left(\re(z)\right),
    \label{eq:sgnz}
\end{align}
where the cut of the square root is chosen along the negative real axis. This choice, although not unique, gives the correct physical result for the overlap Dirac operator in Eq.~\eqref{Dovmu} (see Ref.~\cite{Bloch:2007xi}).

\section{Krylov-Ritz approximations for matrix functions}
\label{Sec:Krylov}

Since we aim at problems with large matrices, as is the case in LQCD, memory and computing power limitations require sophisticated methods to deal with the sign function. For a matrix $A$ of large dimension $n$ the common approach is 
not to compute $f(A)$ but rather its action on a vector, i.e., $y=f(A)x$, which is needed by iterative inverters to compute $f(A)^{-1} b$ or by iterative eigenvalues solvers for $f(A)$. The Krylov-Ritz method approximates the resulting vector in the Krylov subspace
\begin{align}
    \mathcal{K}_{k}(A,x) \equiv \myspan(x,Ax,A^2x,\dotsc,A^{k-1}x)
\end{align}
of $\mathbb{C}^{n}$, implicitly making a polynomial approximation of degree $k-1$ to $f(A)$.
The optimal approximation to $y$ in this subspace is its orthogonal projection $y^\perp_{k}$. 
For $V_{k}=(v_1,\dotsc,v_{k})$, where the $v_i$ form an orthonormal basis of $\mathcal{K}_{k}(A,x)$, an orthogonal projector is given by $P = V_{k} V_{k}^\dagger$,
and we have
\begin{align}
    y &=f(A)x \approx y^\perp_{k} = P f(A)x     .
\end{align}
 However, to compute this projection on the Krylov subspace we already need $y$, which is the quantity we wanted to determine in the first place. Thus, we need to replace this exact projection by an approximation.
To reduce the large dimensionality of the problem one typically projects $A$ on the Krylov subspace using $A_k \equiv P A P$. The projected matrix $A_k$ has dimension $n$ but rank at most $k$. 
The $k$-dimensional image of the projected matrix $A_k$ is defined by the matrix $H_k = V_k^\dagger A V_k$, which is often referred to as Ritz matrix.
The components of $H_k$ are the projection coefficients of $A_k$ in the basis $V_k$, as
 $A_k$ and $H_k$ are related by $A_k = V_k H_k V_k^\dagger$ (in analogy to the vector case).

The Krylov-Ritz approximation \cite{gallopoulos89parallel,saad:209} to $f(A)$ consists in taking the function of the Ritz matrix $H_k$ and lifting it back to the full $n$-dimensional space, 
\begin{align}
f(A) \approx V_k f(H_k) V_k^\dagger .
\label{Ritzapp}
\end{align}
This approximation actually replaces the polynomial interpolating $f$ at the eigenvalues of $A$ by the polynomial interpolating $f$ at the eigenvalues of $H_{k}$, also called Ritz values \cite{saad:209}. Substituting the approximation \eqref{Ritzapp} in $f(A) x$ yields
\begin{align}
    y &\approx V_k f(H_k) V_k^\dagger x = \normx V_{k} f(H_{k}) e_1^{(k)},
    \label{eq:non-nested}
\end{align}
where we choose $v_1$ collinear with $x$, i.e., $v_1 = V_{k}e_1^{(k)} \equiv x/\normx$, with $e_1^{(k)}$ the first unit vector of $\mathbb{C}^k$.
To evaluate the approximation \eqref{eq:non-nested} we do not need to perform the matrix multiplications of Eq.~\eqref{Ritzapp} explicitly.
First, one computes the function $f(H_k)$ of the $k$-dimensional Ritz matrix to high accuracy, using the spectral definition \eqref{fAdef} or a matrix-iterative method. Then, the final approximation is simply a linear combination of the basis vectors $v_i$, with coefficients given by the first column of $f(H_k)$ multiplied with $\normx$.

The Krylov-Ritz approximation described above uses an orthonormal basis of $\mathcal{K}_{k}(A,x)$. For the Hermitian case such a basis can efficiently be constructed using the Lanczos algorithm, which we listed in \ref{alg:lanczos} for completeness. It generates an orthonormal basis and a tridiagonal symmetric $H_k$ using a three-term recurrence relation. The non-Hermitian case is more laborious as the construction of an orthonormal basis is typically performed using the Arnoldi algorithm, which suffers from long recurrences as each basis vector has to be orthogonalized with respect to all the previous ones.
 The two-sided Lanczos algorithm is a suitable alternative \cite{Bloch:2008gh} which uses two three-term recurrences to construct bases $V_{k}=(v_1, \dotsc, v_k)$ and $W_{k}=(w_1, \dotsc w_k)$ of the right, respectively left, Krylov subspaces $\mathcal{K}_{k}(A,x)$ and $\mathcal{K}_{k}(A^\dagger,x)$, which are biorthonormal, i.e., $v_i^\dagger w_j = \delta_{ij}$ (see \ref{alg:bilanczos} for a listing of the algorithm).
The lack of orthogonality of the basis $V_{k}$ prevents the construction of the orthogonal projector needed for the Krylov-Ritz function approximation \eqref{Ritzapp}. 
Nevertheless, the biorthonormality between $V_k$ and $W_{k}$ can be used to construct
an oblique projector $P = V_k W_k^\dagger$ on the right Krylov subspace. The oblique projection of $A$ is $A_k = P A P$ and its $k$-dimensional image is defined by $H_k = W_{k}^\dagger A V_{k}$, which we call two-sided Ritz matrix, such that $A_k = V_k H_k W_k^\dagger$. The matrix $H_k$ generated by the two-sided Lanczos algorithm is tridiagonal.
The two-sided Krylov-Ritz approximation to $f(A)$ then consists in taking the matrix function of $H_k$ and lifting it back to the original space,
\begin{align}
f(A) \approx V_{k} f(H_k) W_k^\dagger .
\end{align}
After applying this approximation of $f(A)$ to $x$ we find an expression which is similar to Eq.~\eqref{eq:non-nested},
\begin{align}
y   &\approx V_{k} f(H_{k}) W_{k}^\dagger x 
    = \normx V_{k} f(H_{k}) e_1^{(k)} ,
    \label{eq:non-nested-oblique}
\end{align}
where the last step assumes that $v_1 = V_{k}e_1^{(k)} \equiv x/\normx $. 
The price paid to achieve short recurrences in the non-Hermitian case is the loss of orthogonality of the projection on the Krylov subspace, which translates in a somewhat lower accuracy of the two-sided Lanczos approximation compared to the Arnoldi approximation, for equal Krylov subspace sizes. Nevertheless, the large gain in speed makes it by far the more efficient method \cite{Bloch:2008gh}.

In the case where $f$ is the sign function, the approximations \eqref{eq:non-nested} and \eqref{eq:non-nested-oblique} require the computation of $\sgn(H_{k})$. Although it could be computed directly with the spectral definition \eqref{fAdef}, matrix-iterative methods are often cheaper for medium sized matrices. We choose to employ the Roberts-Higham iteration (RHi) \cite{Rob80}: Set $S_0 = H_k$ and compute
\begin{align}
    S_{n+1} &= \frac{1}{2}(S_n+S_n^{-1}).
    \label{eq:Roberts}
\end{align}
This iteration converges quadratically to $\sgn(H_k)$, if the sign function for complex arguments is defined by Eq.~\eqref{eq:sgnz}. 
The matrix inversion scales like $k^3$ and so will the RHi. For the QCD application considered here, typically 7 to 10 iterations are necessary to converge within machine precision \cite{Bloch:2007aw,Bloch:2008gh}. 

The hope is that the Krylov-Ritz approximations \eqref{eq:non-nested} and \eqref{eq:non-nested-oblique} are accurate for $k \ll n$. The method is known to work very well as long as no eigenvalues are close to a function discontinuity. However, for the sign function this method suffers from the sign discontinuity along the imaginary axis. If $A$ has eigenvalues close to this discontinuity the approximating polynomial must steeply change from $-1$ to $+1$ over a small interval to give an accurate approximation. This cannot be achieved with a low order polynomial, i.e., the Krylov subspace must be large, which makes the algorithm expensive. The common solution to this problem is to use deflation, where the contribution of the eigencomponents associated to these critical eigenvalues to the sign function is computed exactly.\footnote{In practice we deflated the eigenvalues with smallest modulus $|\lambda|$ instead of those with smallest absolute real part $|\re\lambda|$, as the former are more efficiently determined numerically, and both choices yield almost identical deflations for the operator $\gamma_5\Dw(\mu)$ of Eq.~\eqref{Dovmu}. The reason for this is that, as long as the chemical potential $\mu$ is not unusually large, the spectrum looks like a very narrow bow-tie shaped strip along the real axis, and the sets of eigenvalues with smallest absolute real parts and smallest magnitudes will nearly coincide. In the following we therefore define the deflation gap $\Delta$ as the largest deflated eigenvalue in magnitude, i.e., $\Delta = \max|\lambda_\text{defl}|$.} The Krylov subspace approximation is then performed in a deflated space, i.e., the subspace where the directions along the critical eigenvectors have been removed. We refer to the literature for details \cite{Bloch:2007aw}.

\begin{figure}
\includegraphics[width=0.49\textwidth]{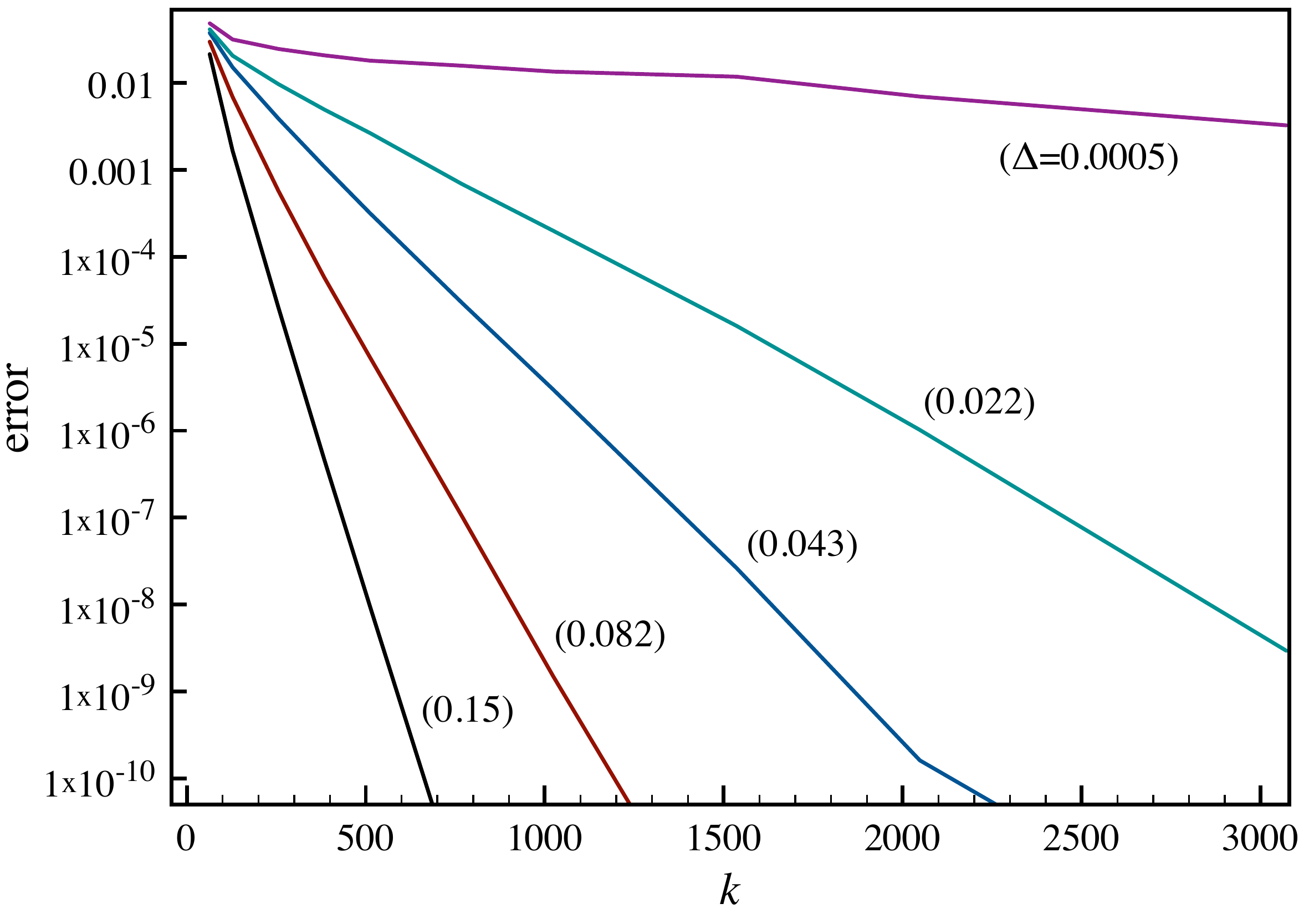}
\hfill
\includegraphics[width=0.49\textwidth,type=pdf,ext=.pdf,read=.pdf]{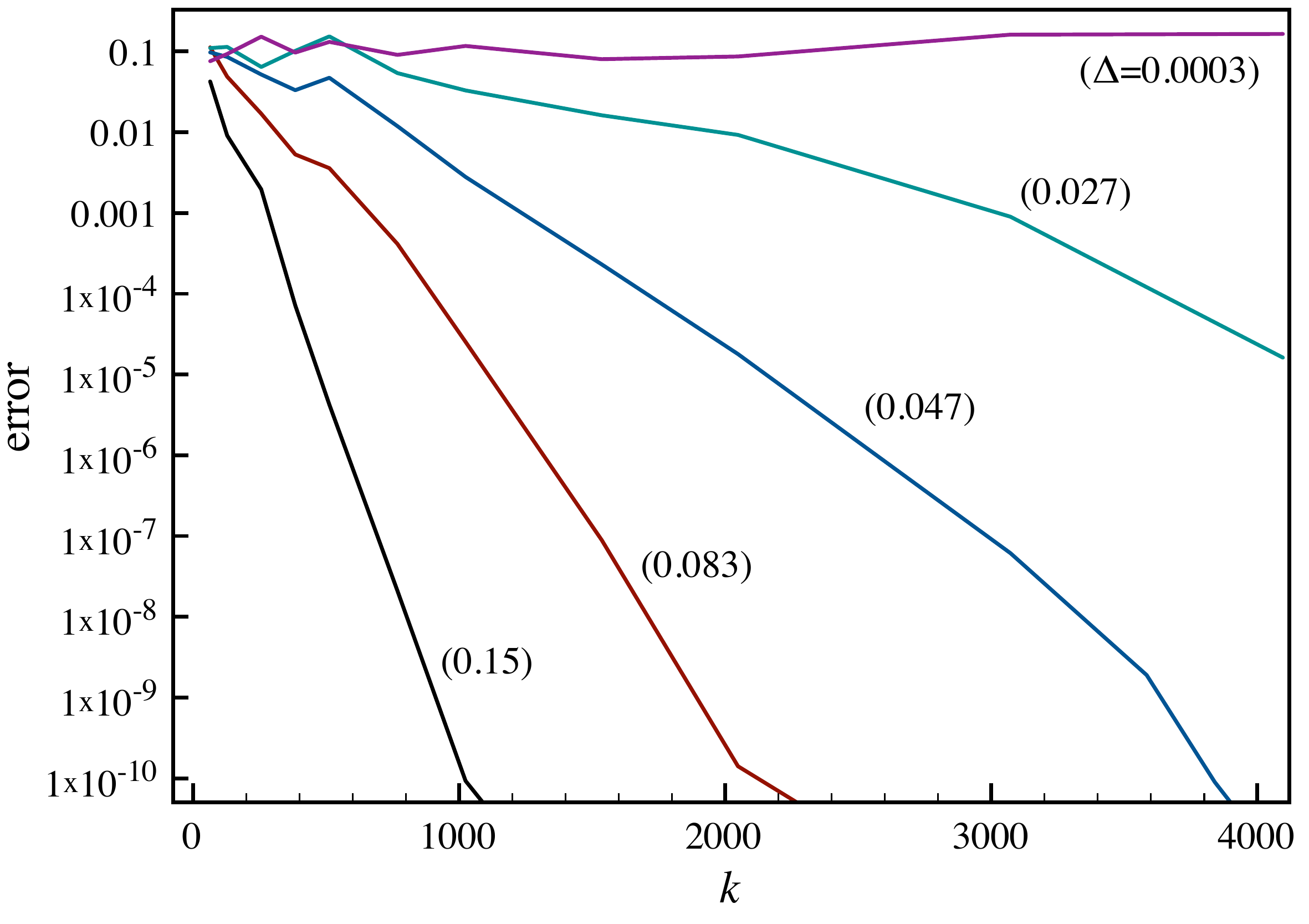}
\caption{Accuracy of the Krylov subspace approximation for $y=\sgn(A)x$, where $A$ is $\gamma_5 \Dw(\mu)$ for a $6^4$ lattice (for a lattice volume $V$ the matrix $\gamma_5 \Dw$ has dimension $12 V$, such that $\dim(A) = 15552$ here). Left pane: Hermitian case ($\mu=0$) using the Lanczos method, 
right pane: non-Hermitian case with chemical potential $\mu=0.3$ using the two-sided Lanczos method. 
The relative error $\varepsilon$ is shown as a function of the Krylov subspace size $k$ for different deflation gaps $\Delta$ (given in parenthesis). }
\label{fig:6666_LS}
\end{figure}

The convergence of the Krylov-Ritz approximations to the matrix sign function is 
illustrated in Fig.~\ref{fig:6666_LS}: the Lanczos approximation for the Hermitian case on the left, and the two-sided Lanczos approximation for the non-Hermitian case on the right. The accuracy of the approximation cannot be determined by comparing to the exact value $\sgn(A) x$, as its evaluation by direct methods is too costly if A is large. To obtain an estimate for the error, we compute $\tilde x \approx \sgn(A)^2 x$ (by applying the Krylov-Ritz approximation twice in succession), which should equal $x$ if the approximation to the sign function were exact, and then take $\varepsilon=|\tilde x-x|/2\normx$ as a measure for the error. This error estimate proved to be consistent with the true error obtained by comparing the approximation to the exact solution for $4^4$ and $6^4$ lattices, and will therefore be used for all lattice sizes. Here, and in all subsequent tests,  we choose the test vector $x=(1,\ldots,1)$.
As expected, the accuracy improves with increasing Krylov subspace size $k$, and a larger deflation gap $\Delta$, corresponding to a higher number of deflated eigenvectors, leads to a faster convergence. 
For a given accuracy and equal deflation gap, the subspace size $k$ required for non-Hermitian $A$ is larger than for Hermitian $A$. 

\begin{figure}
\includegraphics[width=0.49\textwidth]{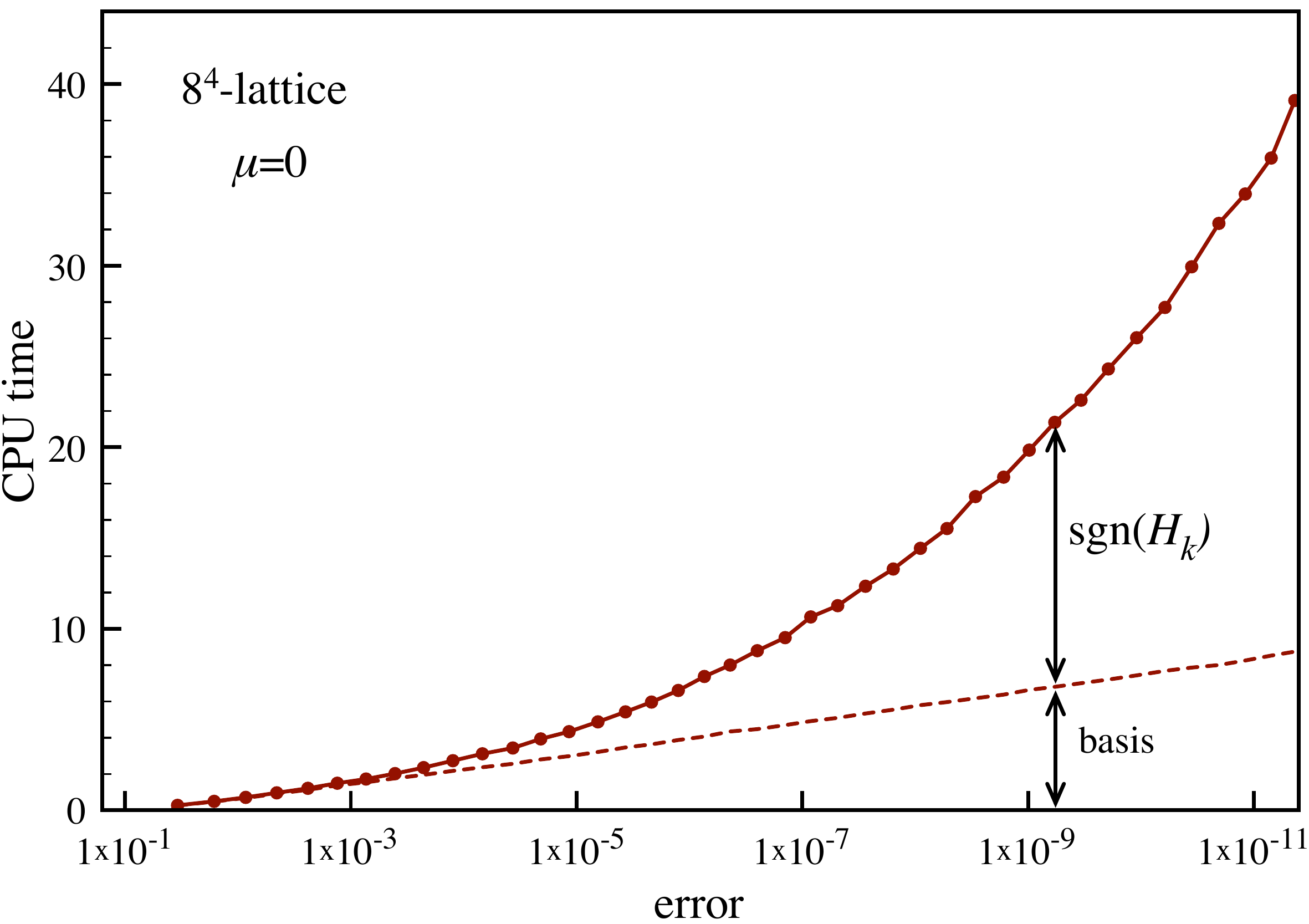}
\hfill
\includegraphics[width=0.49\textwidth,type=pdf,ext=.pdf,read=.pdf]{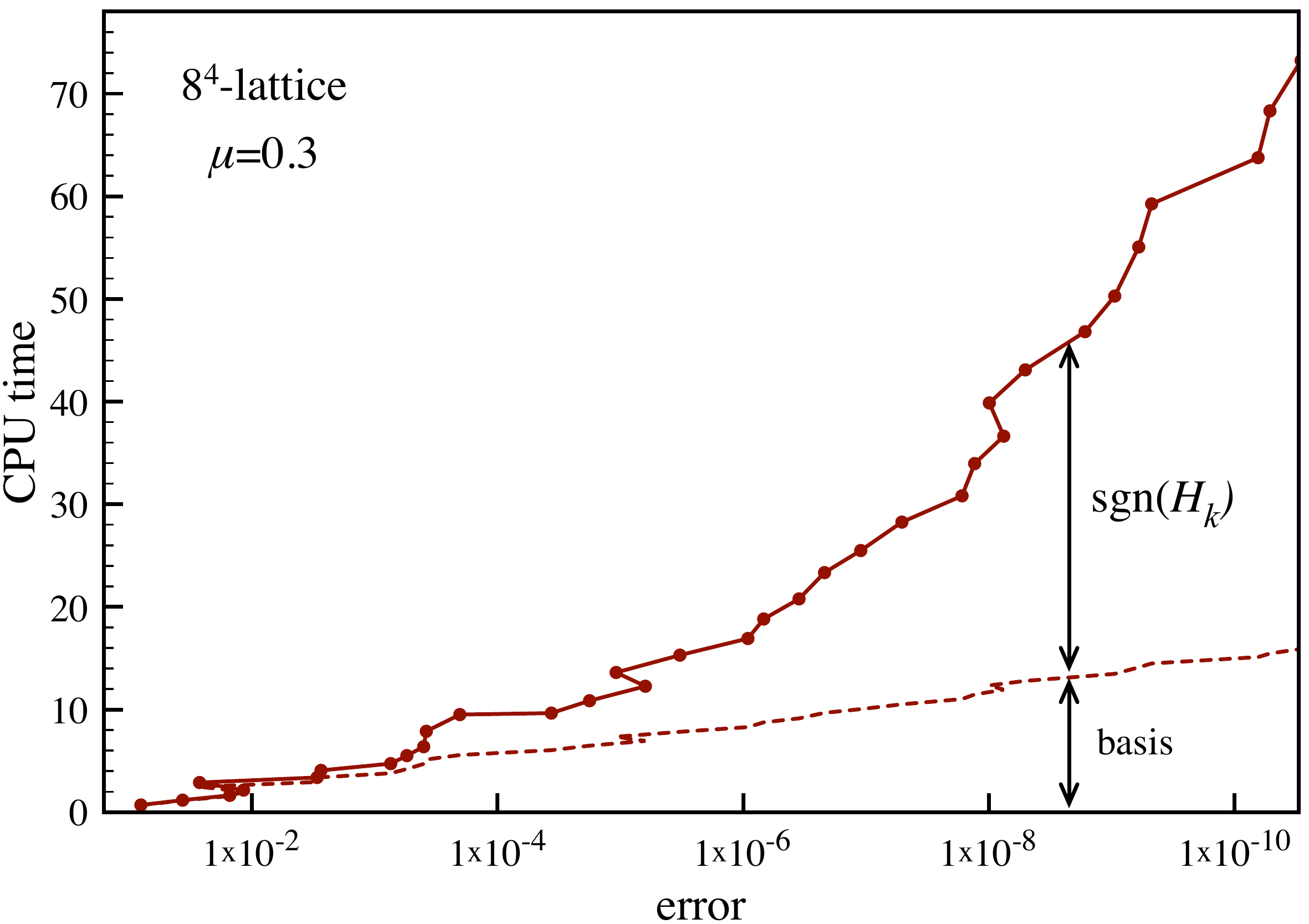}
\caption{
CPU time $t$ (in seconds) versus accuracy for an $8^4$ lattice configuration in the Hermitian case with deflation gap $\Delta = 0.055$ (left) and the non-Hermitian case with $\mu = 0.3$ and deflation gap $\Delta = 0.107$ (right). The full line shows the total time required to compute $\sgn(A)x$, while the dashed line gives the time needed to construct the Krylov basis. The difference between both lines represents the time taken by the RHi to compute $\sgn(H_{k})$. The irregular convergence pattern for the non-Hermitian case is a well-known feature of the two-sided Lanczos algorithm.}
\label{fig:8888_runtime}
\end{figure}

To analyze the efficiency of the algorithm we briefly sketch the three major contributions to the total CPU time.
For each matrix $A$ the deflation requires the computation of the critical eigenvalues and the corresponding eigenvectors.  The time needed by the rest of the algorithm strongly depends on the eigenvalue gap, as the Krylov subspace size can be reduced if the deflation gap is increased. As mentioned at the beginning of this section, the product $f(A) x$ is usually needed for many source vectors $x$, e.g., as part of an iterative inversion. In this case the expensive deflation of $A$ only needs to be performed once in an initialization step, while the Krylov subspace part of the algorithm will be repeated for each new vector $x$. For this reason 
we assume from now on that an initial deflation has been performed and we will concentrate on the efficiency of the Krylov subspace part of the algorithm.
We discern two main components in the Krylov-Ritz method: the construction of the Krylov basis using the Lanczos or two-sided Lanczos algorithms, where the computation time grows linearly with the subspace size $k$, and the RHi to compute $\sgn(H_{k})$, which scales as $k^3$. 
Figure \ref{fig:8888_runtime} illustrates these last two contributions. 
For high accuracy the Krylov subspace becomes large such that the cost of the RHi dominates the total CPU time of the Krylov-Ritz approximation and the method becomes too costly.
In the following, the implementation of the Krylov-Ritz approximation for which $\sgn(H_{k})$ is computed using Eq.~\eqref{eq:Roberts} will be referred to as \textit{non-nested method}. 
In the next section we will present a \textit{nested} Krylov subspace method, which drastically reduces the cost to compute $\sgn(H_{\kout}) e_1^{(\kout)}$ and vastly improves the overall efficiency of the Krylov-Ritz approximation.

\section{Nested Krylov subspace method for the sign function}
\label{Sec:nested}

\subsection{Nesting and preconditioning}

We introduce a new method which speeds up the expensive computation of the vector $\sgn(H_{k})e_1^{(k)}$ required in the Krylov-Ritz approximations \eqref{eq:non-nested} and \eqref{eq:non-nested-oblique} to $\sgn(A) x$. 
The idea is to approximate this matrix-vector product by a further Krylov-Ritz approximation, using a second, nested Krylov subspace  (specified below) of size $\kin \ll \kout$, i.e.,
\begin{align}
\sgn(H_{k})e_1^{(k)} = V_{\kin} \sgn(H_{\kin}) e_1^{(\kin)} ,
\label{innersign}
\end{align}
where $V_{\kin}$ is the matrix containing the basis vectors of the inner Krylov subspace, constructed with the Lanczos or two-sided Lanczos method, and $H_{\kin}$ is the inner Ritz or two-sided Ritz matrix. The $\sgn(H_{\kin})$ is computed using the RHi on the inner Ritz matrix $H_{\kin}$.
After substituting this result in Eq.~\eqref{eq:non-nested} and \eqref{eq:non-nested-oblique} we get the nested approximation
\begin{align}
y   &\approx \normx \, V_{\kout} V_{\kin} \sgn(H_{\kin}) e_1^{(\kin)}
\label{nested}
\end{align}
 to $\sgn(A) x$. 
By introducing an additional Krylov subspace, the number of operations necessary to compute $\sgn(H_{\kout})e_1^{(k)}$ is reduced from ${\cal O}(\kout^3)$  in the non-nested method to $\order(\kin^3)$ + $\order(\kout \kin)$.
If  $\kin \ll \kout$ this will very much improve the efficiency of the Krylov-Ritz approximation.

The obvious choice for the inner Krylov subspace is $\mathcal{K}_{\kin}(H_{\kout},e_1^{(\kout)})$. However, it is easy to see that approximations in this Krylov subspace will not improve the efficiency of the method. 
The Ritz matrix $H_{\kin}$ of the Krylov subspace $\mathcal{K}_{\kin}(H_{\kout},e_1^{(\kout)})$ will only contain information coming from the $\kin\times \kin$ upper left corner of $H_{\kout}$, because of the tridiagonal nature of $H_{\kout}$ and the sparseness of the source vector $e_1^{(\kout)}$. 
This will effectively cut down the size of the outer Krylov subspace from $\kout$ to $\kin$,
which will substantially worsen the accuracy of the approximation if $\kin$ is chosen much smaller than $\kout$.
Nonetheless, the nested Krylov subspace method can be made to work efficiently
if we perform an initial preconditioning step on the tridiagonal Ritz matrix, replacing\footnote{The factor $1/2$ is chosen for convenience. For $p=1$ the transformation actually mimics the first step of the RHi \eqref{eq:Roberts}. }
\begin{align}
H_k \to  H_k' = \frac12 \left[p H_{\kout}+ (p H_{\kout})^{-1} \right],
\label{precond}
\end{align}
with $p$ a positive real number, and construct the approximation to $\sgn(H_{\kout})e_1^{(\kout)}$ in the Krylov subspace $\mathcal{K}_{\kin}(H_{\kout}',e_1^{(\kout)})$.  
This alternate Krylov subspace can be used to compute $\sgn(H_{\kout})e_1^{(\kout)}$ because the transformation leaves the sign unchanged. 
To show this, we note that both matrices have identical eigenvectors, as a matrix and its inverse share the same eigenvectors, and that the sign of their eigenvalues satisfies
\begin{align}
\sgn\frac12\left(p z+\frac{1}{pz}\right) &= \sgn\re\left(p z+\frac{1}{p z}\right) = \sgn\re\left(p z+\frac{p z^*}{|p z|^2}\right)  
    = \sgn\Bigg[\underbrace{\left(1+\frac{1}{|pz|^2}\right)}_{>0}\re(p z)\Bigg] = \sgn \left(z\right),
\label{proofprec}
\end{align}
where we used the definition \eqref{eq:sgnz}. Hence, $\sgn(H_{\kout}')=\sgn(H_{\kout})$ according to Eq.~\eqref{fAdef}.\footnote{If $H_k$ is not diagonalizable, the equality can be shown by applying Eq.~\eqref{proofprec} to the integration variable in the integral representation \eqref{eq:matfun}.}
\begin{figure}
\centering
\includegraphics[width=0.49\textwidth]{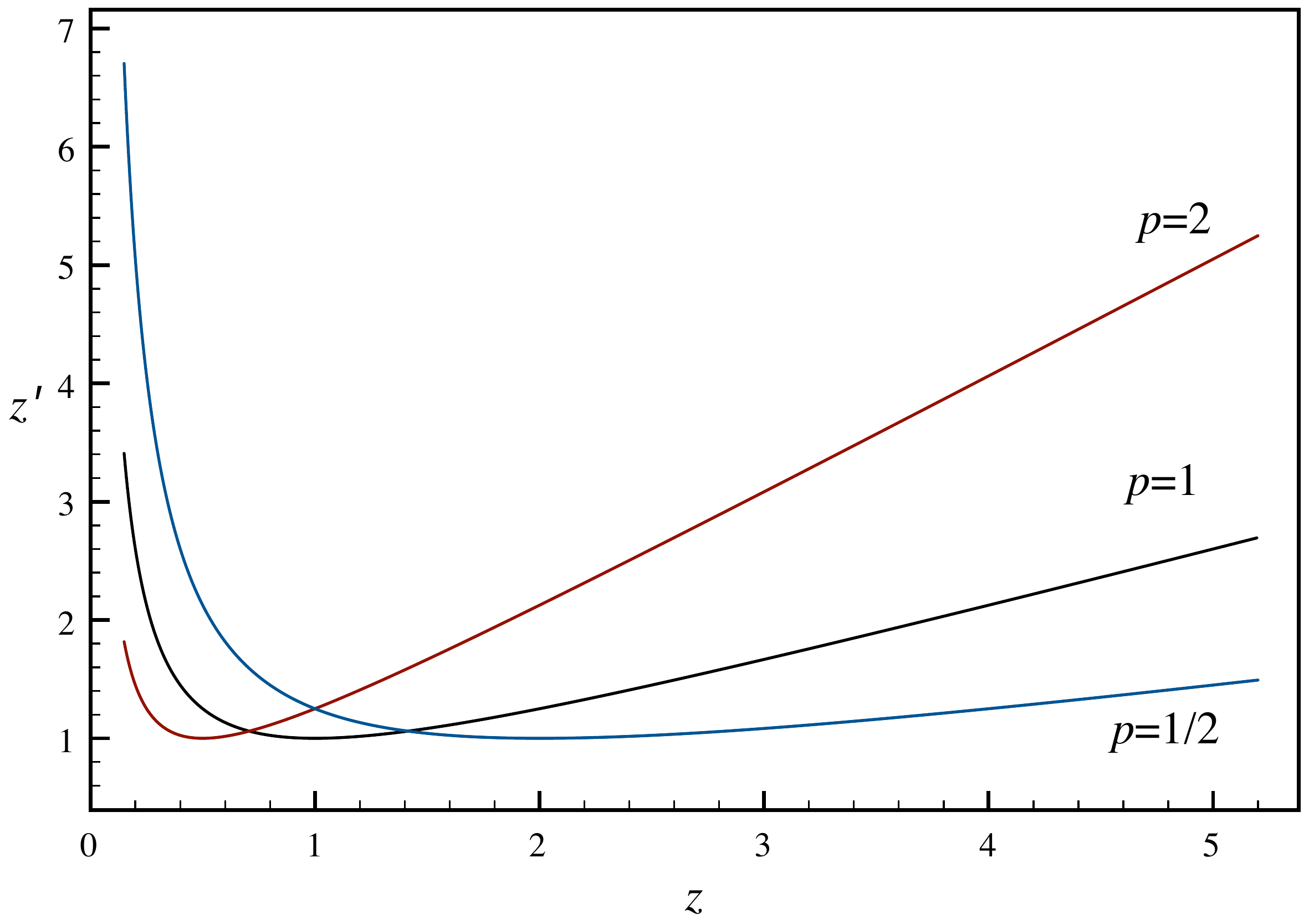}
\caption{ Mapping of the preconditioning step $z'=(pz+1/pz)/2$ for positive real eigenvalues and various values of $p$.}
\label{fig:precond}
\end{figure}

As $H_\kout$  is tridiagonal the cost of its inversion, required in \eqref{precond}, is only of ${\cal O}(\kout)$.
Moreover, as the transformation increases the relative gap between the spectrum and the singularity along the imaginary axis (see below), we expect a clear gain in efficiency for the inner Krylov-Ritz approximation, characterized by  $\kin \ll \kout$. 

\begin{figure}
\centering
\includegraphics[width=0.6\textwidth]{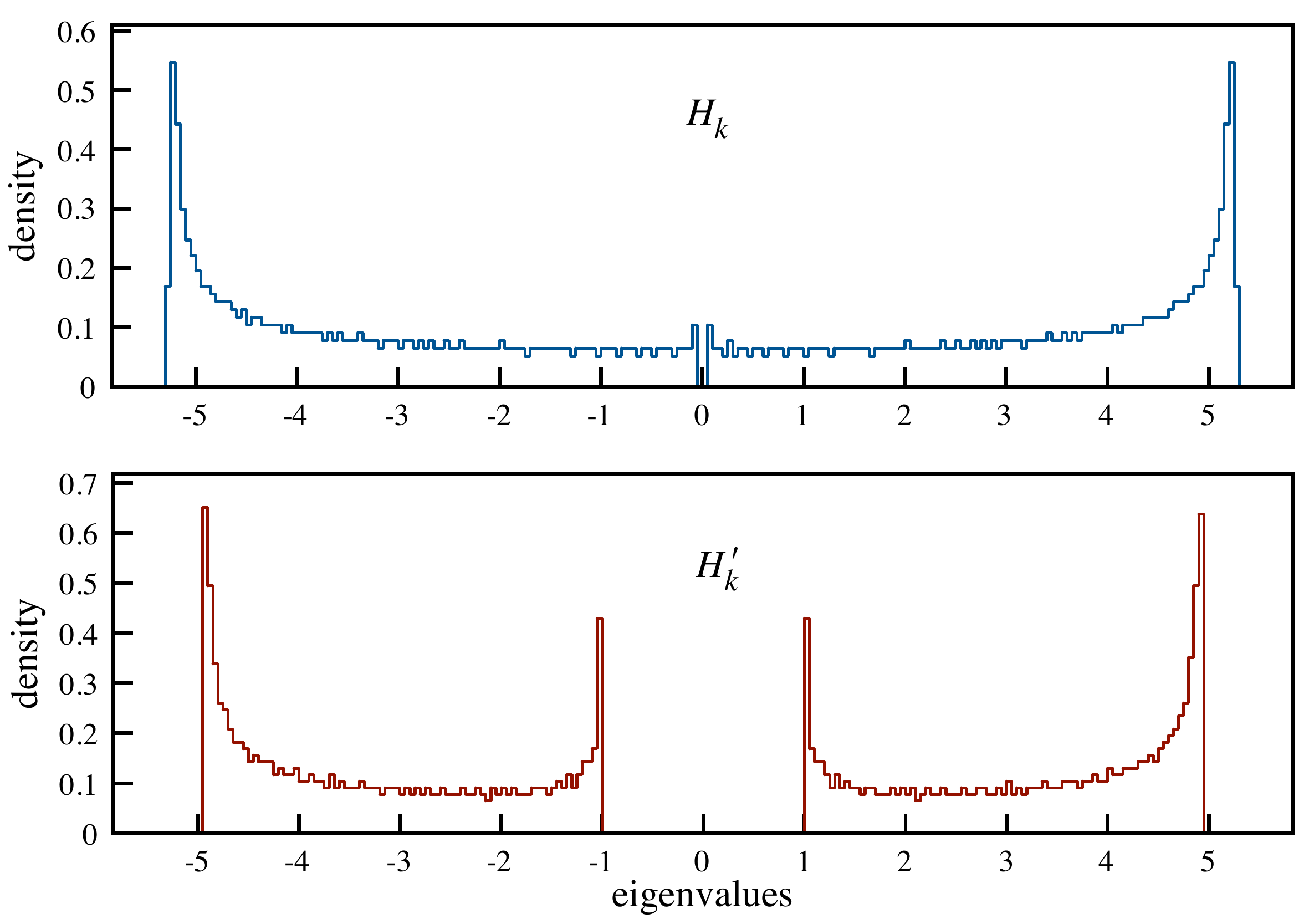}
\caption{Upper pane: density of eigenvalues of $H_{\kout}$ in the Hermitian case for an $8^4$ lattice with $\kout=1536$. The spectrum has a narrow deflation gap $\Delta=0.055$. The optimal $p$-factor \eqref{popt} for the transformation \eqref{precond} is $p_\text{opt} \approx 1.86$ (using $z_{\min}=\Delta$ and $z_{\max} = 5.26$).
The lower pane shows the corresponding eigenvalue density of the transformed matrix $H_{\kout}'$, where 
the condition number is improved by a factor  ${\cal F} = 19.3$ (see Eq.~\eqref{condimp}).
 }
\label{fig:eval_density}
\end{figure}

For a Hermitian matrix the transformation induced by the preconditioning step is illustrated in Fig.~\ref{fig:precond} for real positive eigenvalues (for negative values the graph would be reflected with respect to the origin). 
The factor $p$ is chosen to optimize the effect of the transformation on the relative distance to the imaginary axis, which in the Hermitian case corresponds to a minimization of the condition number. 
We examine the condition number for the Hermitian case, assuming that the spectral support of $H_k$ is similar to that of the original matrix $A$, after deflation.
As can be seen from Fig.~\ref{fig:precond}, after transformation the smallest eigenvalue (in absolute value) is $z'_{\min}=1$, while the largest will be given by the transform of either the smallest or largest eigenvalues of $H_k$. The smallest condition number will be achieved when both values are identical, i.e., for $p$ satisfying\footnote{In practice $p_\text{opt}$ is only known approximately, as it is computed from spectral information of A instead of $H_{\kout}$. However, 
this has no significant impact on the performance of the nested method.}
\begin{align}
\frac{1}{2}\left(p z_{\min} + \frac{1}{p z_{\min}} \right) \stackrel{!}{=} 
\frac{1}{2}\left(p z_{\max} + \frac{1}{p z_{\max}} \right) 
\quad \Rightarrow \quad p_\text{opt} = \sqrt{\frac{1}{z_{\min}z_{\max}}} ,
\label{popt}
\end{align}
where $z_{\min} = \min |z|$ and $z_{\max} = \max |z|$, for $z$ in the spectrum of $H_{\kout}$, 
and the largest transformed eigenvalue will be
\begin{align}
z'_{\max} = \frac12\left(\sqrt{\frac{z_{\max}}{z_{\min}}} + \sqrt{\frac{z_{\min}}{z_{\max}}}\right) \approx
\frac12 \sqrt{\frac{z_{\max}}{z_{\min}}} .
\end{align}
In the Hermitian case, the transformation \eqref{precond} therefore reduces the condition number $C$ by a factor 
\begin{align}
{\cal F} = \frac{C}{C'} = \frac{z_{\max}}{z_{\min}} \Bigg/
\frac12\left(\sqrt{\frac{z_{\max}}{z_{\min}}} + \sqrt{\frac{z_{\min}}{z_{\max}}}\right)
\approx 2\sqrt{\frac{z_{\max}}{z_{\min}}} .
\label{condimp}
\end{align}
The effect of the preconditioning of the Ritz matrix for a typical spectrum of $\gamma_5 \Dw$ in lattice QCD is illustrated in Fig.~\ref{fig:eval_density} for the Hermitian case. 
The top and bottom graphs depict the spectra of $H_{\kout}$ and $H_{\kout}'$, respectively.
The spectrum of the original Ritz matrix has only a small gap at zero, while the gap for  the transformed matrix is large. In this example, the condition number is almost improved by a factor 20.
In general, the value of $z_{\max}$ for $\gamma_5 \Dw$ varies only slightly with the choice of the simulation parameters and ${\cal F}$ will mainly depend on the deflation gap.

\begin{figure}
\centering
\begin{minipage}{0.35\textwidth}
\includegraphics[width=\textwidth]{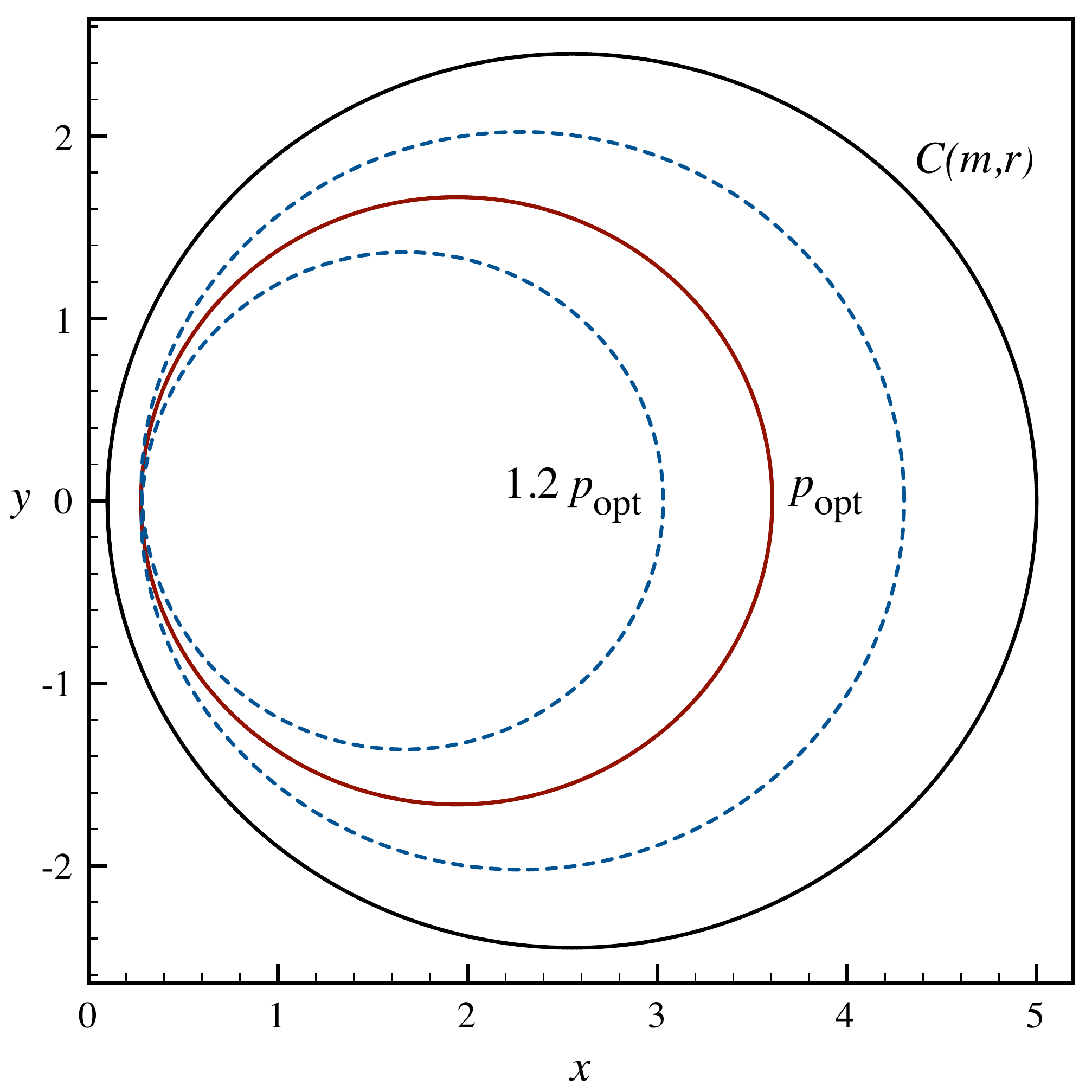}
\end{minipage}\hspace{10mm}
\begin{minipage}{0.45\textwidth}
\vspace{-4mm}
\includegraphics[width=\textwidth]{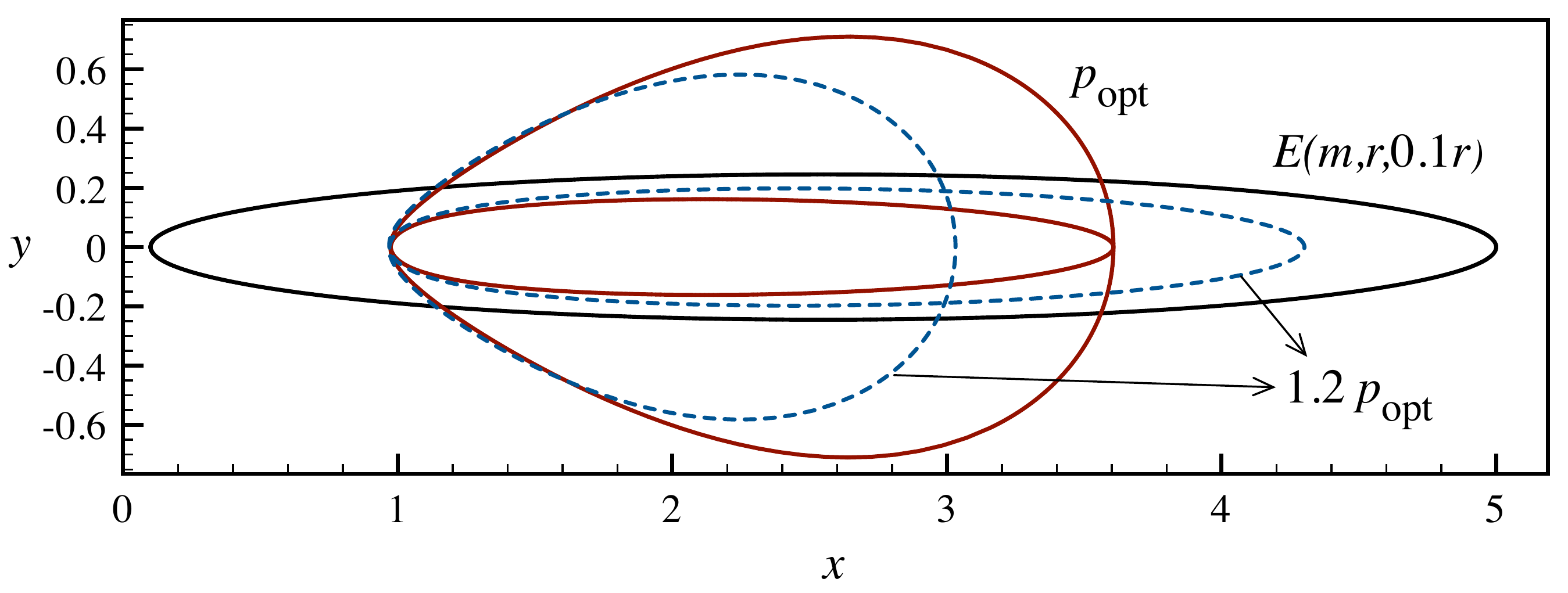}
\caption{ Mapping of the transformation $z'=(pz+1/pz)/2$ for complex $z=x+iy$ on a circle $C(m,r)$ with center $m=2.55$ and radius $r=2.45$ (left) and for z on the ellipse $E(m,r,0.1\, r)$ (right). In both plots the black curve shows the original $z$-values, the red curve the transformed values $z'$ with optimal $p$, and the blue dashed lines the transformed values with a sub-optimal $p$.}
\label{fig:precond_complex}
\end{minipage}
\end{figure}

For the non-Hermitian case, let us assume that the complex spectrum is contained in the circles $C(-m,r) \cup C(m,r)$, with real center $m> 0$ and radius $r<m$.
The optimal $p$, maximizing the relative distance from the imaginary axis for the transformed spectrum, is  still given by Eq.~\eqref{popt} which now simplifies to $p_\text{opt}=(m^2-r^2)^{-1/2}$. For this choice the transformed eigenvalues are contained in the circles $C(-m',r') \cup C(m',r')$ with center $m'=(m_s+1/m_s)/2$ and radius $r'=(m_s-1/m_s)/2$, where $m_s \equiv p_\text{opt}m$. This is illustrated in the left panel of Fig.~\ref{fig:precond_complex}, where we show the transformation of a circle $C(m,r)$ for the optimal and a sub-optimal value of $p$. For sub-optimal $p$ the transformation yields an inner and an outer circle-like contour, which merge into the circle $C(m',r')$ when $p \to p_\text{opt}$. For $p_\text{opt}$ the relative distance from the imaginary axis will be maximal and we expect the transformation \eqref{precond} to work best. The gain in efficiency will however not be as large as for the Hermitian case. This can be quantified by the relative distance to the imaginary axis, in analogy to the calculation performed above for the Hermitian case. For the original spectrum we define the relative distance as
\begin{align}
d &\equiv \frac{\min|\re z|}{\max|\re z|} = \frac{m-r}{m+r}
\intertext{and for the transformed spectrum}
d' &\equiv \frac{\min|\re z'|}{\max|\re z'|} = \frac{m'-r'}{m'+r'} = \frac{1}{m_s^2} 
= \frac{m^2-r^2}{m^2} .
\end{align}
The improvement factor due to the transformation is given by the ratio of these distances, yielding
\begin{align}
{\cal F} = \frac{\;d'}{\!d}  = \left(\frac{m+r}{m}\right)^2  = \left(2 - \frac{\Delta}{m}\right)^2,
\label{condimpNH}
\end{align}
where we wrote $r=m-\Delta$, with $\Delta$ the deflation gap. When $\Delta \ll m$ we will have ${\cal F} \approx 4$. For the example shown in the left plot of Fig.~\ref{fig:precond_complex} the transformation generates an improvement by a factor ${\cal F} = 3.84$, as computed with Eq.~\eqref{condimpNH}. 
 
\begin{figure}
\centering
\includegraphics[width=0.5\textwidth,type=pdf,ext=.pdf,read=.pdf]{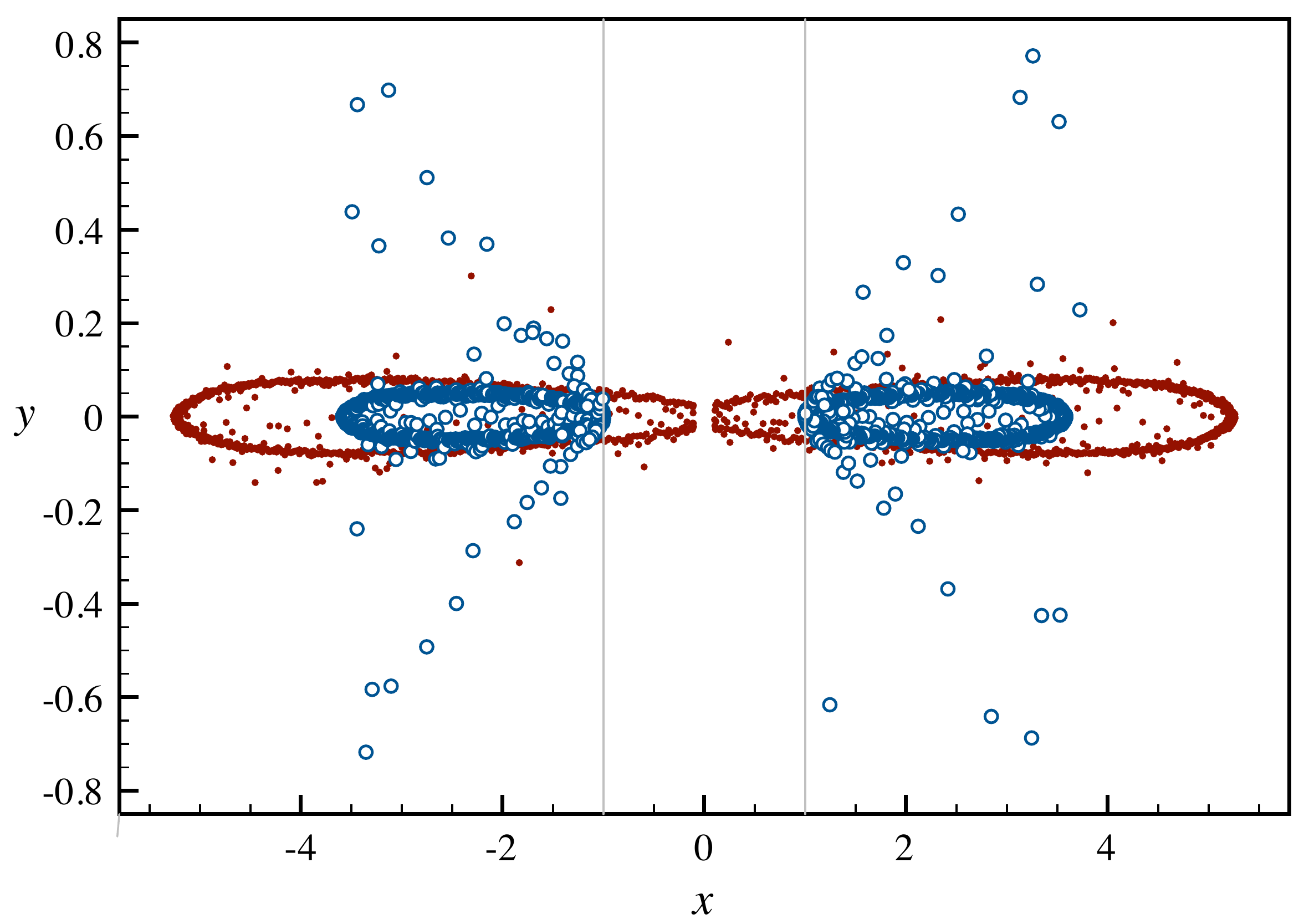}
\caption{Red dots: spectrum of $H_{\kout}$ in the non-Hermitian case for an $8^4$ lattice with $\kout=1580$, $\mu=0.3$ and deflation gap $\Delta=0.107$. The optimal $p$-factor \eqref{popt} for the transformation \eqref{precond} is $p_\text{opt} \approx 1.335$ (using $z_{\min}=\Delta$ and $z_{\max} = 5.243$).
Blue circles: the corresponding spectrum of the transformed matrix $H_{\kout}'$. As desired, the transformed eigenvalues are well away from the imaginary axis (the vertical lines at $x=\pm 1$ serve to guide the eye). Note the different scales on the $x$ and $y$ axes.
 }
\label{fig:complex_eval_density}
\end{figure}

In lattice QCD at nonzero baryon density $\gamma_5\Dw$ is usually weakly non-Hermitian and, after deflation, the spectra are contained in ellipses $E(-m,a,b) \cup E(+m,a,b)$, with center $m \in \mathbb{R}^+$ and major and minor axes $a$ and $b$ along the real and imaginary axes, respectively. The transformation \eqref{precond} of an ellipse $E(m,a,b)$  with aspect ratio $a/b=10$ is illustrated in the right panel of Fig.~\ref{fig:precond_complex}. For such a narrow ellipse the transformed spectrum is qualitatively similar to the Hermitian case, as all the eigenvalues are transformed to the right of $z' = 1$, i.e., away from the imaginary axis, such that the high efficiency of the transformation is still guaranteed. The optimal value $p_\text{opt}$ is again determined by \eqref{popt} with $p_\text{opt} = (m^2-a^2)^{-1/2}$, as it maximizes the relative distance from the imaginary axis.
The transformation is illustrated for a realistic test case of lattice QCD in Fig.~\ref{fig:complex_eval_density}, where the eigenvalues and transformed eigenvalues of the Ritz matrix for $\gamma_5 \Dw(\mu)$ are shown for $\mu=0.3$.

As we will see below the preconditioning step significantly speeds up the Krylov-Ritz approximation in its application to lattice QCD at zero and nonzero chemical potential.

\subsection{Convergence}
\label{Sec:convergence}

In this section we investigate the convergence properties of the nested method.
The method was implemented to compute the sign function of $\gamma_5 \Dw(\mu)$ needed by the overlap operator \eqref{Dovmu}, for both the Hermitian and the non-Hermitian case.  
Whenever the matrix has eigenvalues close to the imaginary axis, these critical eigenvalues are first deflated to open up a deflation gap, necessary to keep the Krylov subspace within a reasonable size (see Sec.~\eqref{Sec:Krylov}).
Our implementation uses Chroma \cite{Edwards:2004sx} to compute the Wilson operator. The linear algebra is performed with BLAS and LAPACK routines. 
To ensure the efficiency of the nested method a judicious implementation of the preconditioning step \eqref{precond}, used to construct the inner Krylov subspace, is needed.
Explicitly inverting the tridiagonal matrix $H_{\kout}$ to form the full matrix $H_{\kout}'$, then constructing the basis of the inner Krylov subspace by  successive full matrix-vector multiplications would make a rather inefficient algorithm.
To construct the inner Krylov subspace we do not need to construct the full matrix $H_{\kout}'$ explicitly, but only have to apply $H_{\kout}'$ to $\kin-1$ vectors of $\mathbb{C}^{\kout}$ (in the non-Hermitian case ${H_{\kout}'}^\dagger$ is also needed). 
These products are best computed using the $LU$ decomposition of $H_{\kout}$, which is ${\cal O}(k)$ and thus especially efficient for tridiagonal matrices.
A detailed listing of the algorithm is given in \ref{alg:nested}.

\begin{figure}
\includegraphics[width=0.49\textwidth]{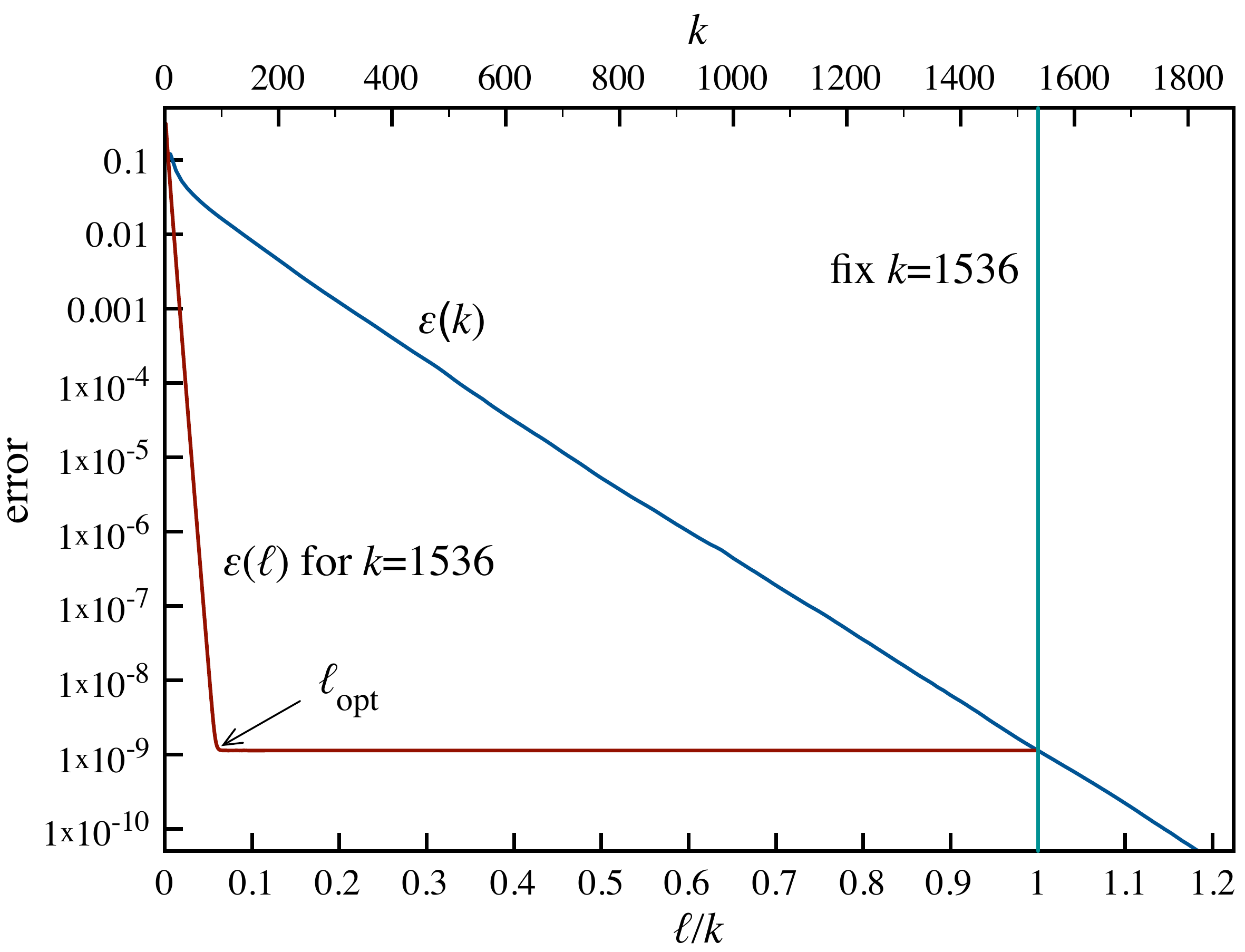}
\hfill
\includegraphics[width=0.49\textwidth,type=pdf,ext=.pdf,read=.pdf]{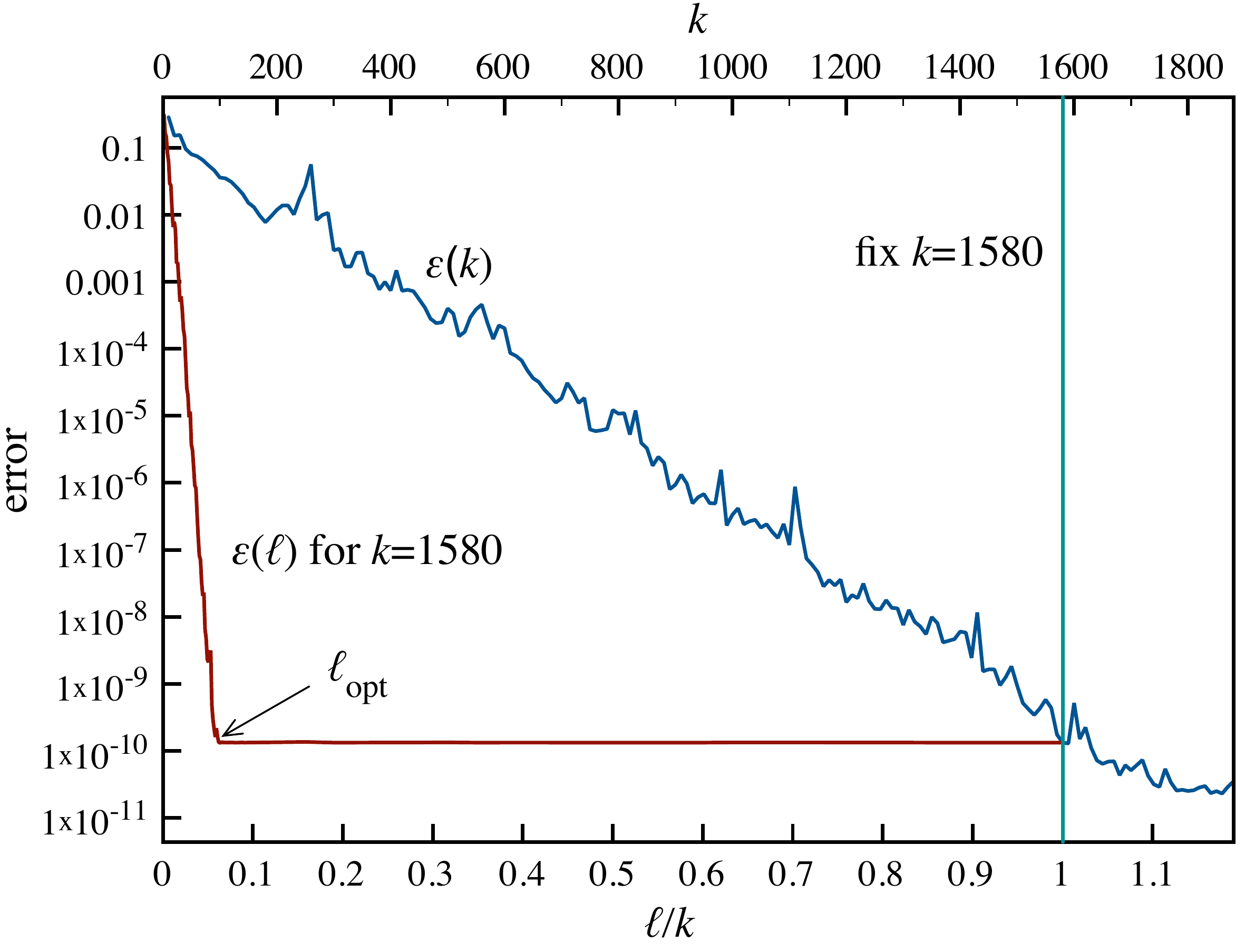}
\caption{Accuracy $\varepsilon$ of the nested method for an $8^4$ lattice configuration. Hermitian case with deflation gap $\Delta=0.055$ (left) and non-Hermitian case with $\mu=0.3$ and deflation gap $\Delta=0.107$ (right). $\varepsilon(\kout)$ shows how the error of the non-nested method decreases with growing Krylov subspace (blue line). The vertical line fixes the size $\kout$ of the outer Krylov space used in the nested method. $\varepsilon(\kin)$ shows the accuracy of the nested method, for fixed $\kout$, as a function of the size $\kin$ of the inner Krylov subspace (red line). The rapid convergence illustrates the efficiency of the nested method. 
The smallest value of $\kin$ for which optimal convergence is reached is denoted by $\kin_\text{opt}$.
Note that we always restrict ourselves to even Krylov subspace sizes, as odd values systematically give a somewhat worse accuracy because of spurious near-zero eigenvalues occurring in the Ritz matrix. 
}
\label{fig:8888_k1_dependence}
\end{figure}

\begin{figure}
\includegraphics[width=0.49\textwidth]{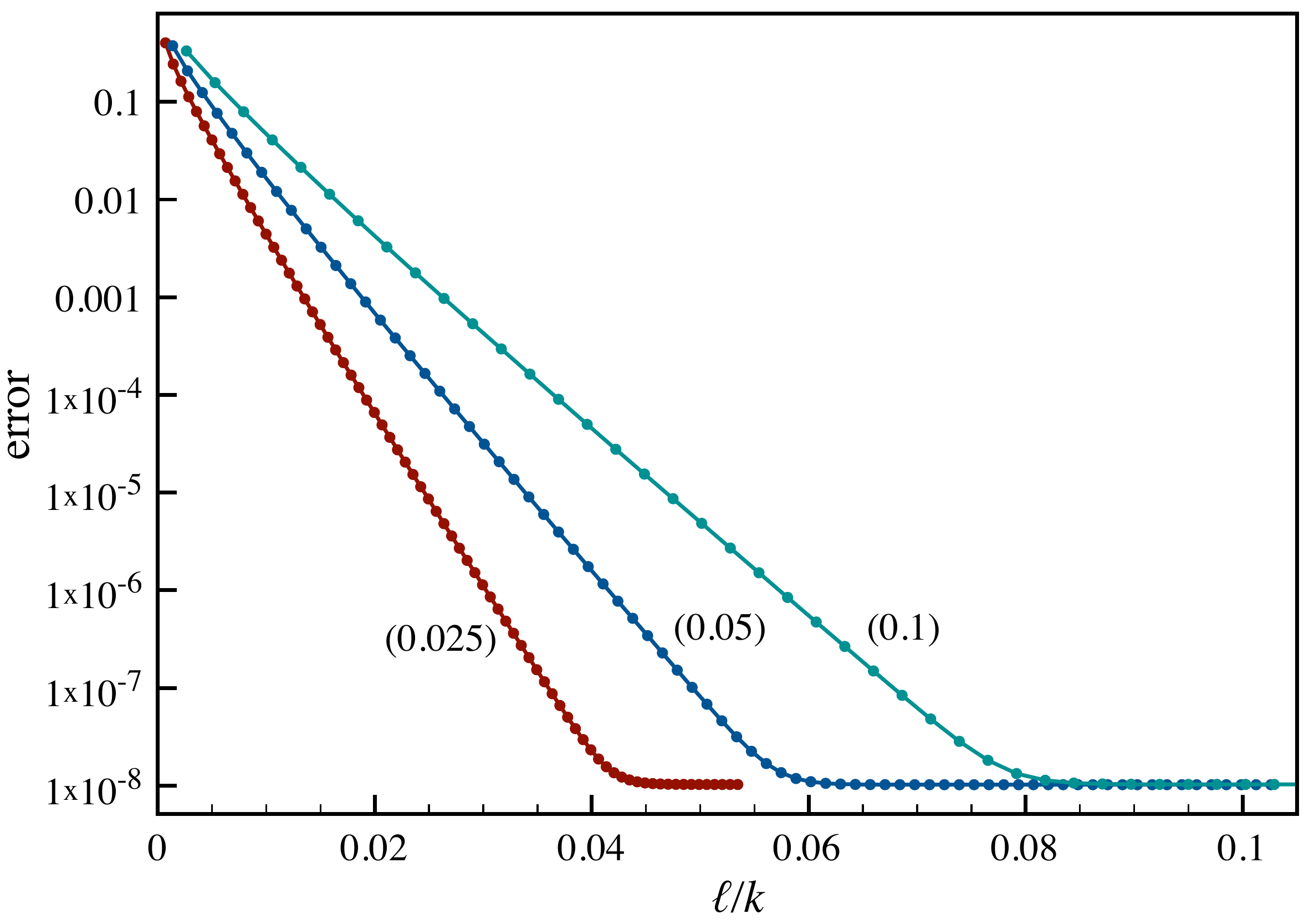}
\hfill
\includegraphics[width=0.49\textwidth,type=pdf,ext=.pdf,read=.pdf]{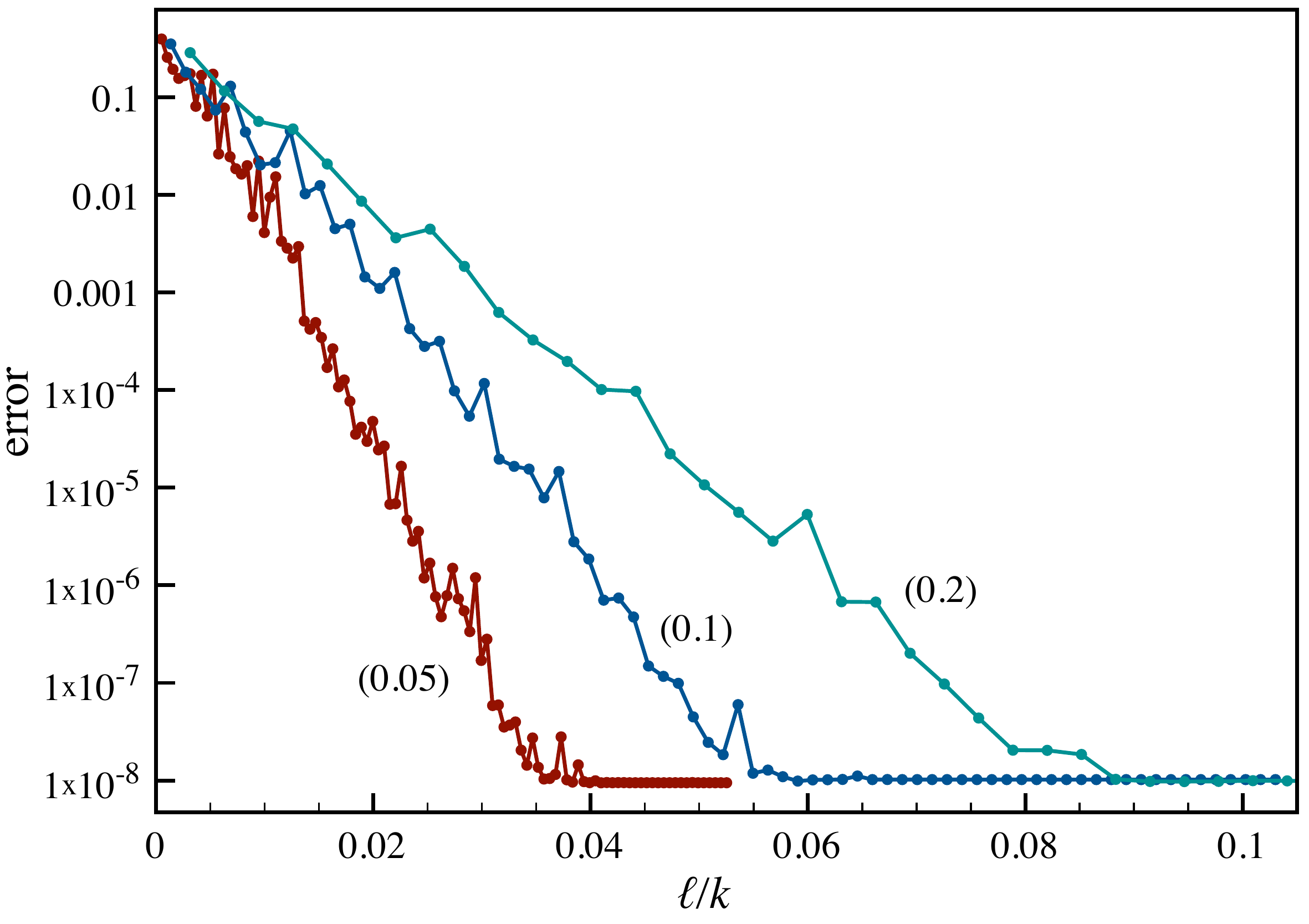}
\caption{Convergence of the nested method for an $8^4$ lattice configuration as a function of the relative inner Krylov subspace size $\kin/\kout$, for various deflation gaps (given in parenthesis). For each gap the value of $\kout$ is chosen such that an accuracy of $10^{-8}$ is achieved. Left: Hermitian case with $\kout=2806, 1462 \text{ and }758$ for deflation gap $\Delta=0.025,0.05\text{ and }0.1$. Right: non-Hermitian case with $\mu=0.3$ and $\kout=3808,1456\text{ and }634$ for $\Delta=0.05, 0.1\text{ and }0.2$. Again, the irregular convergence pattern for the non-Hermitian case is characteristic for the two-sided Lanczos algorithm.
}
\label{fig:8888_k1_defl_dependence}
\end{figure}

The overall accuracy of the nested approximation \eqref{nested} depends on the parameters $\kout$ and $\kin$, defining the sizes of the outer and inner Krylov subspaces, respectively. 
For $\kin \to \kout$  the solution of the nested method will converge to that of the non-nested method with Krylov subspace size $\kout$ and accuracy $\varepsilon_k$, so its total error will also converge to $\varepsilon_k$.
To investigate the accuracy of the nested algorithm, our strategy is to fix the outer Krylov subspace size $\kout$, corresponding to a certain desired accuracy, and vary the inner Krylov subspace size $\kin$. We show the convergence results for an $8^4$ lattice configuration in Fig.~\ref{fig:8888_k1_dependence}, for both the Hermitian and non-Hermitian case. As expected the nested method reaches the accuracy of the non-nested method when its size is large enough. 
Surprisingly however, this happens for $\kin \ll \kout$, as the convergence of the inner Ritz approximation seems to be extremely fast. 
The smallest value of $\kin$ for which optimal convergence is reached will be called $\kin_\text{opt}$. 
The fast convergence is closely related to the large improvement in condition number discussed in the previous section. 
We also showed in Eq.~\eqref{condimp} how the improvement of the condition number, due to the preconditioning of the Ritz matrix $H_{\kout}$, depends on the deflation gap. A smaller gap will yield a larger improvement, and vice-versa. This in turn will influence the convergence rate of the nested method. 
Figure~\ref{fig:8888_k1_defl_dependence} verifies that the result $\lopt \ll \kout$ remains valid for different deflation gaps.
The figure also illustrates that the somewhat larger reduction in condition number achieved for a smaller gap yields an accordingly smaller ratio $\kin_\text{opt}/\kout$ (approximately proportional to the ratio of the respective improvement factors ${\cal F}$). 
This is an additional advantage as the size reduction is largest when the outer subspace is large.
 In all cases, the inner Krylov subspace can be taken much smaller than the outer subspace, such that the efficiency of the Krylov-Ritz method is substantially boosted, as will be shown in the benchmarks below.
 
We also verified that the convergence curves are fairly insensitive to the choice of the source vector and lattice configuration.
The fast convergence property of the nested method is generic, regardless of the simulation details, for both the Hermitian and non-Hermitian case, 
even though the precise value of $\kin_\text{opt}$ depends on the lattice size, the simulation parameters, the deflation gap and the desired overall accuracy (determined by $\kout$).

\subsection{Benchmarks}
\label{benchmarks}

\begin{figure}
\includegraphics[width=0.49\textwidth]{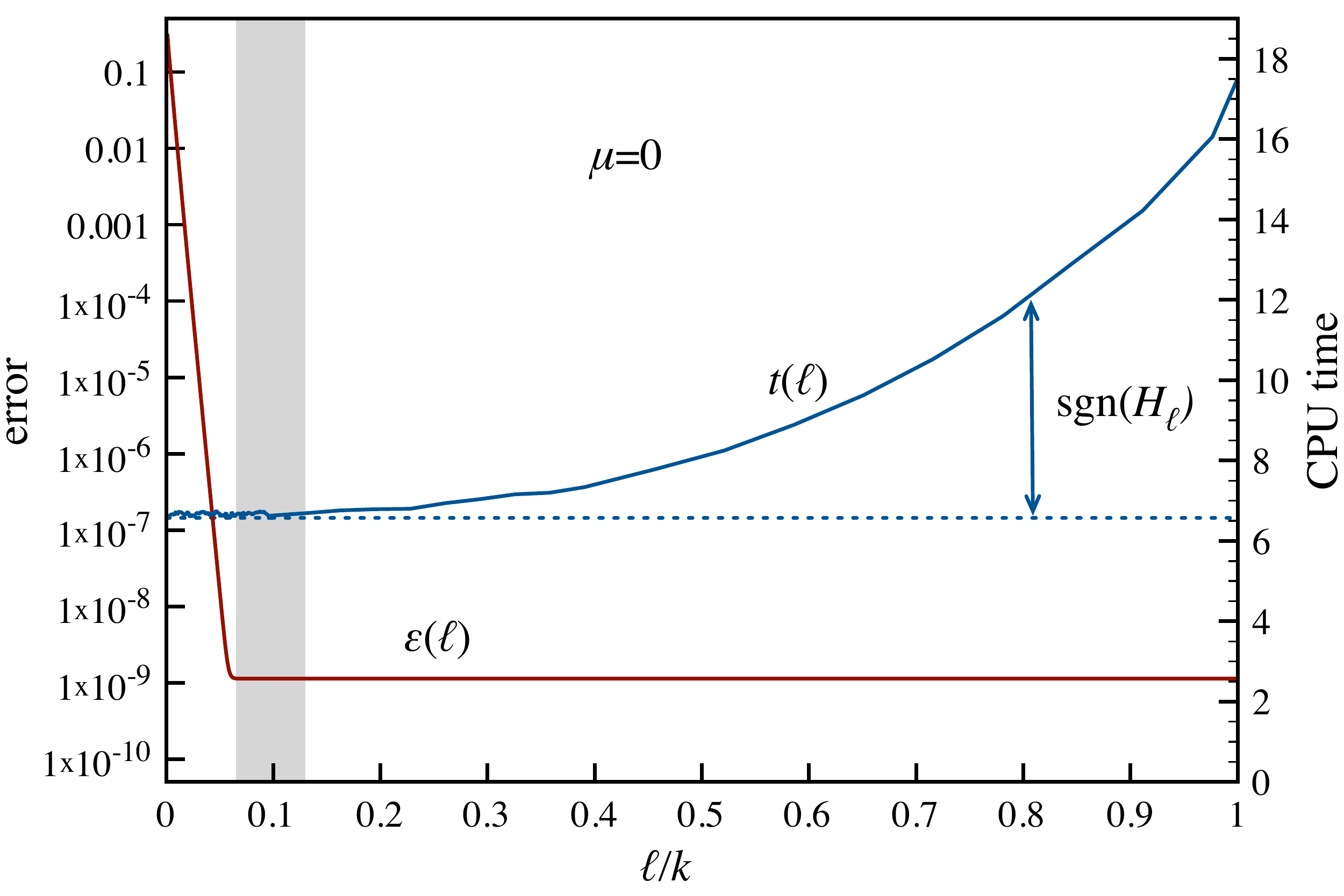}
\hfill
\includegraphics[width=0.49\textwidth,type=pdf,ext=.pdf,read=.pdf]{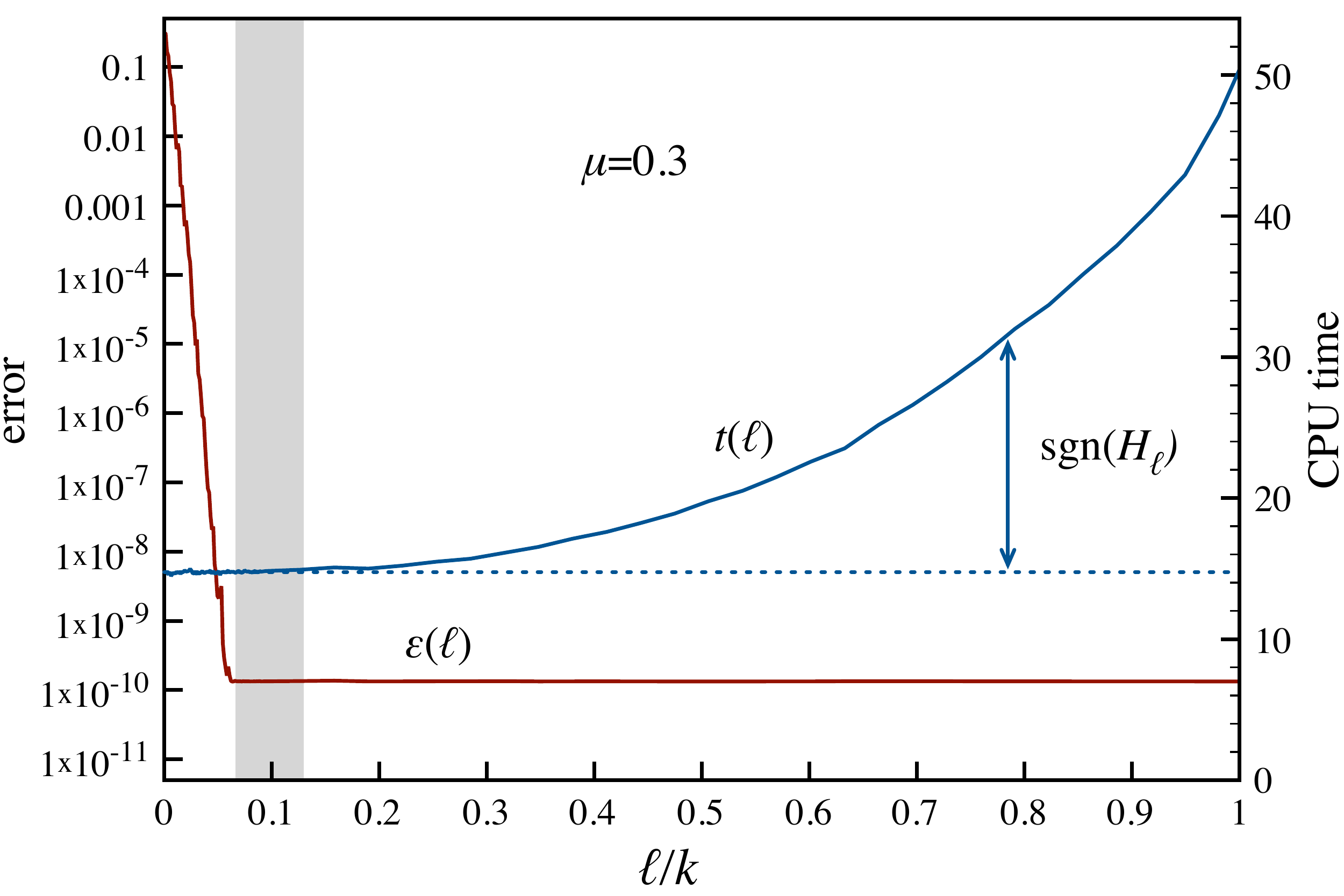}
\caption{Error and CPU usage of the nested method for lattice size $8^4$. Hermitian case with deflation gap $\Delta=0.055$ and fixed $\kout = 1536$ (left), and non-Hermitian case with $\mu=0.3$, $\Delta=0.107$ and $\kout=1580$ (right). $\varepsilon(\kin)$ shows the accuracy versus inner Krylov subspace size $\kin$ (red line). $t(\kin)$ shows the total CPU time in seconds (solid blue line), while the horizontal dashed line measures the time needed to construct the basis in the outer Krylov subspace. The difference between both lines corresponds to the time taken by the RHi to compute $\sgn(H_{\kin})$.
The vertical band highlights the operational window of the nested method, i.e., the region in $\kin$ where the accuracy is optimal, but the CPU-time used to compute $\sgn(H_\kin)$ is negligible. 
}
\label{fig:8888_k1_time_dependence}
\end{figure}

With the fast convergence ($\lopt \ll \kout$) discussed in the previous section, we can expect a substantial gain in computation time when using the nested method. The total CPU time consumed by the nested method is illustrated in Fig.~\ref{fig:8888_k1_time_dependence} for the Hermitian case (left) and the non-Hermitian case (right). The size of the outer Krylov subspace is kept fixed, such that its construction gives a constant contribution to the run time, depicted by the horizontal dashed line. The contribution to the CPU time which varies with $\kin$ mainly comes from the computation of $\sgn(H_{\kin})$ with the RHi and is proportional to $\kin^3$. For $\kin\approx \kout$ the total run time of the nested method is about equal to that of the non-nested method. However, as illustrated by the $\varepsilon(\kin)$ curve (red line) and discussed in Sec.~\ref{Sec:convergence}, $\kin$ can be chosen much smaller while preserving the accuracy of the non-nested method. 
The central result, illustrated by the vertical band in Fig.~\ref{fig:8888_k1_time_dependence}, is that there exists an interval in $\kin$ for which the accuracy is still optimal, but the CPU time needed to compute $\sgn(H_\kin)$ with the RHi is negligible compared to the time required to construct the outer Krylov subspace.
There is therefore no need to make a compromise between run time and accuracy, as both can be optimized simultaneously. 
The error in this range is the minimal error achievable with the given size of the outer Krylov subspace, while the run time is completely dominated by the cost for building the basis in that subspace. 
The nested method is able to quench the CPU time needed for the computation of $\sgn(H_{\kout}) e_1^{(\kout)}$ without affecting the accuracy of the Krylov-Ritz approximation.

\begin{figure}
\setlength{\figwidth}{0.42\textwidth}
\centering
\includegraphics[width=\figwidth,type=pdf,ext=.pdf,read=.pdf]{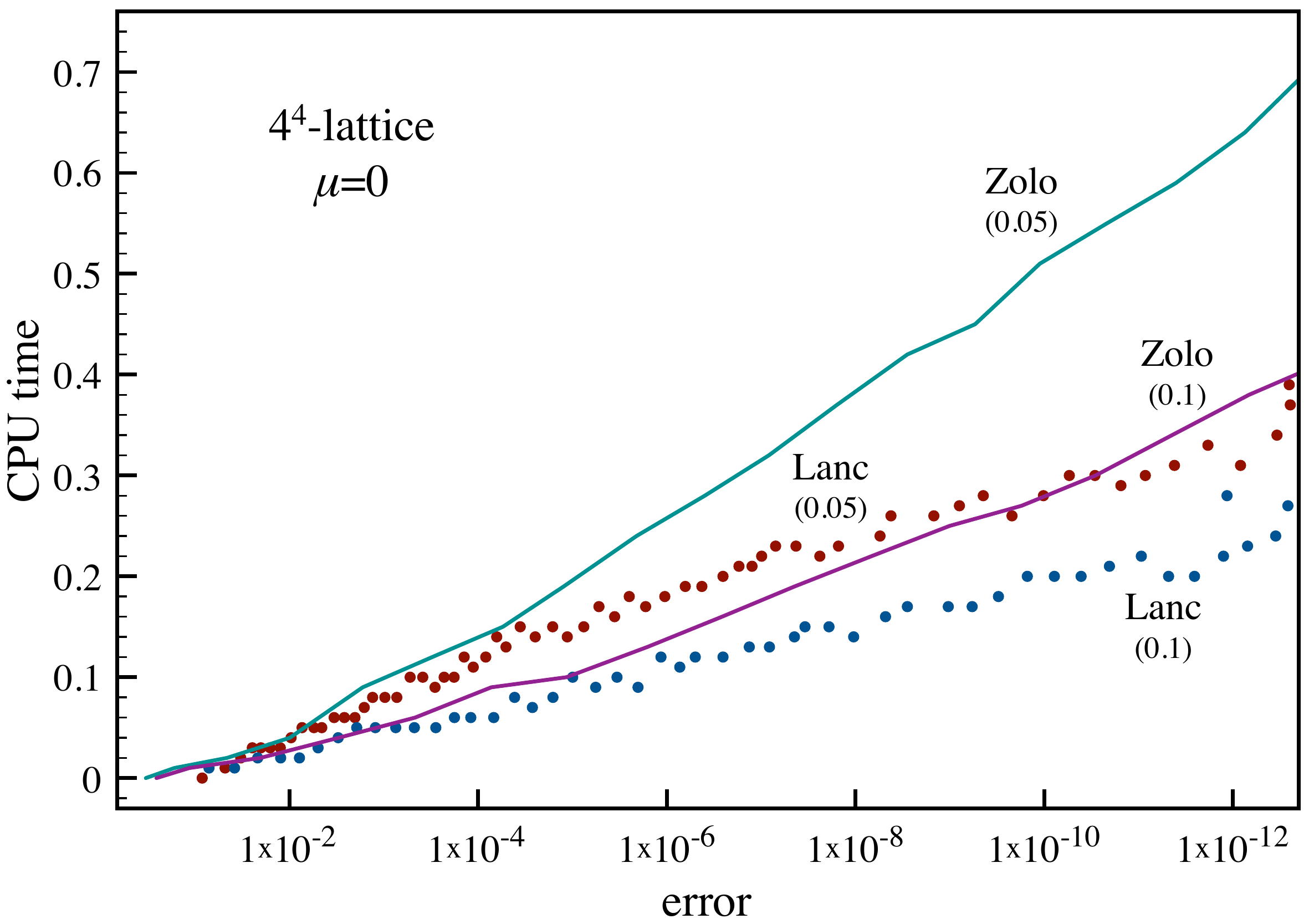}
\hspace{10mm}
\includegraphics[width=\figwidth,type=pdf,ext=.pdf,read=.pdf]{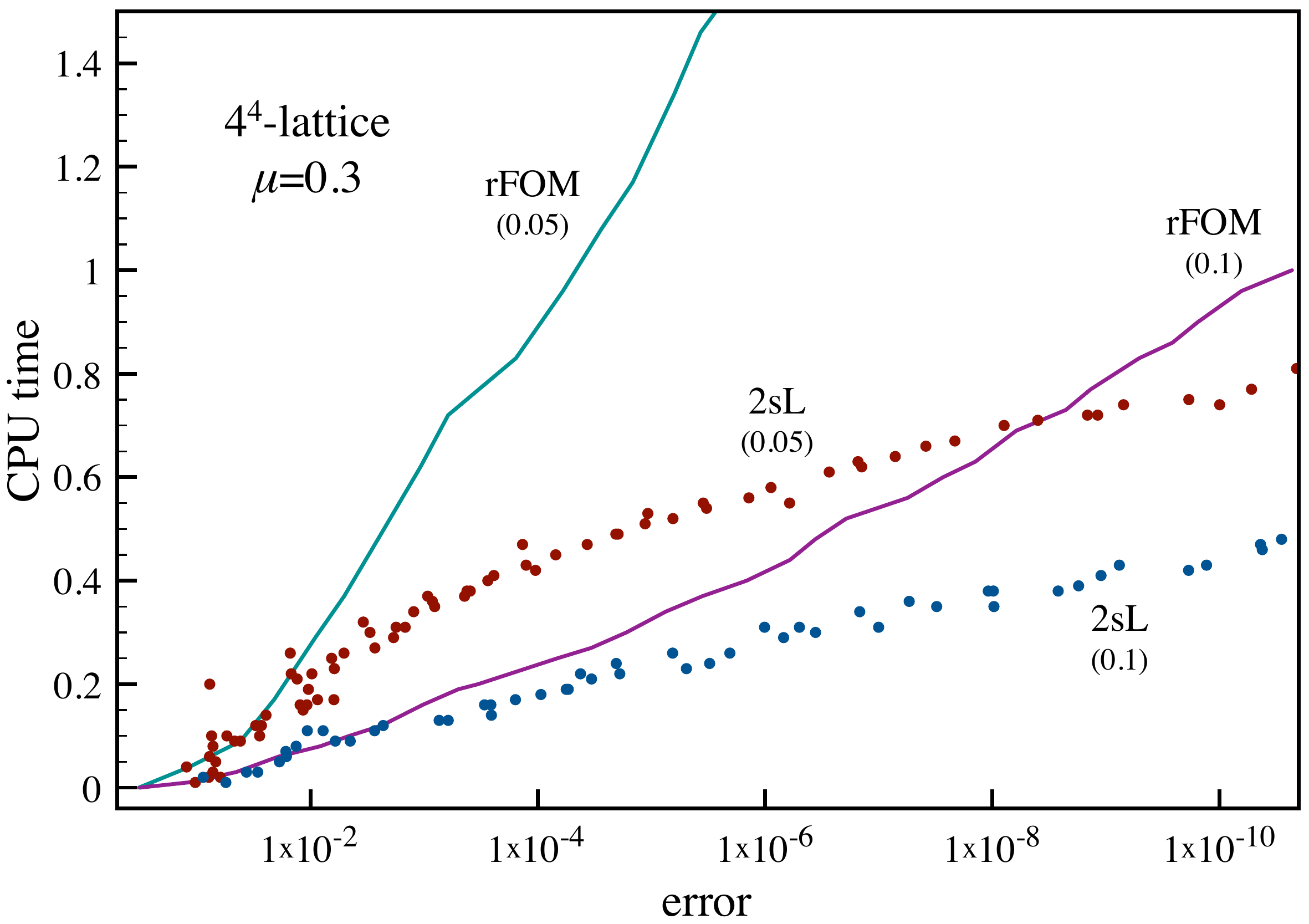}
\includegraphics[width=\figwidth,type=pdf,ext=.pdf,read=.pdf]{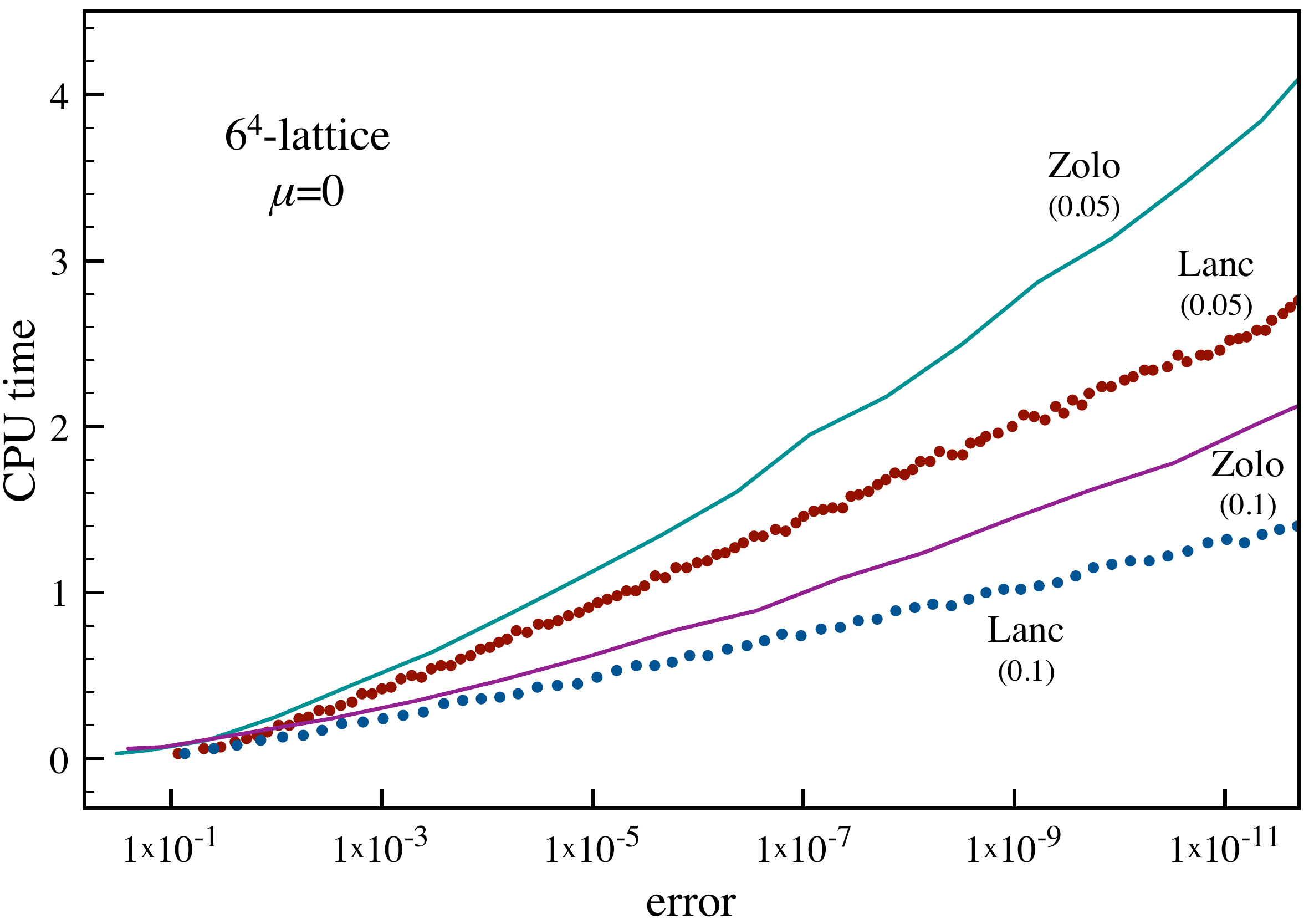}
\hspace{10mm}
\includegraphics[width=\figwidth,type=pdf,ext=.pdf,read=.pdf]{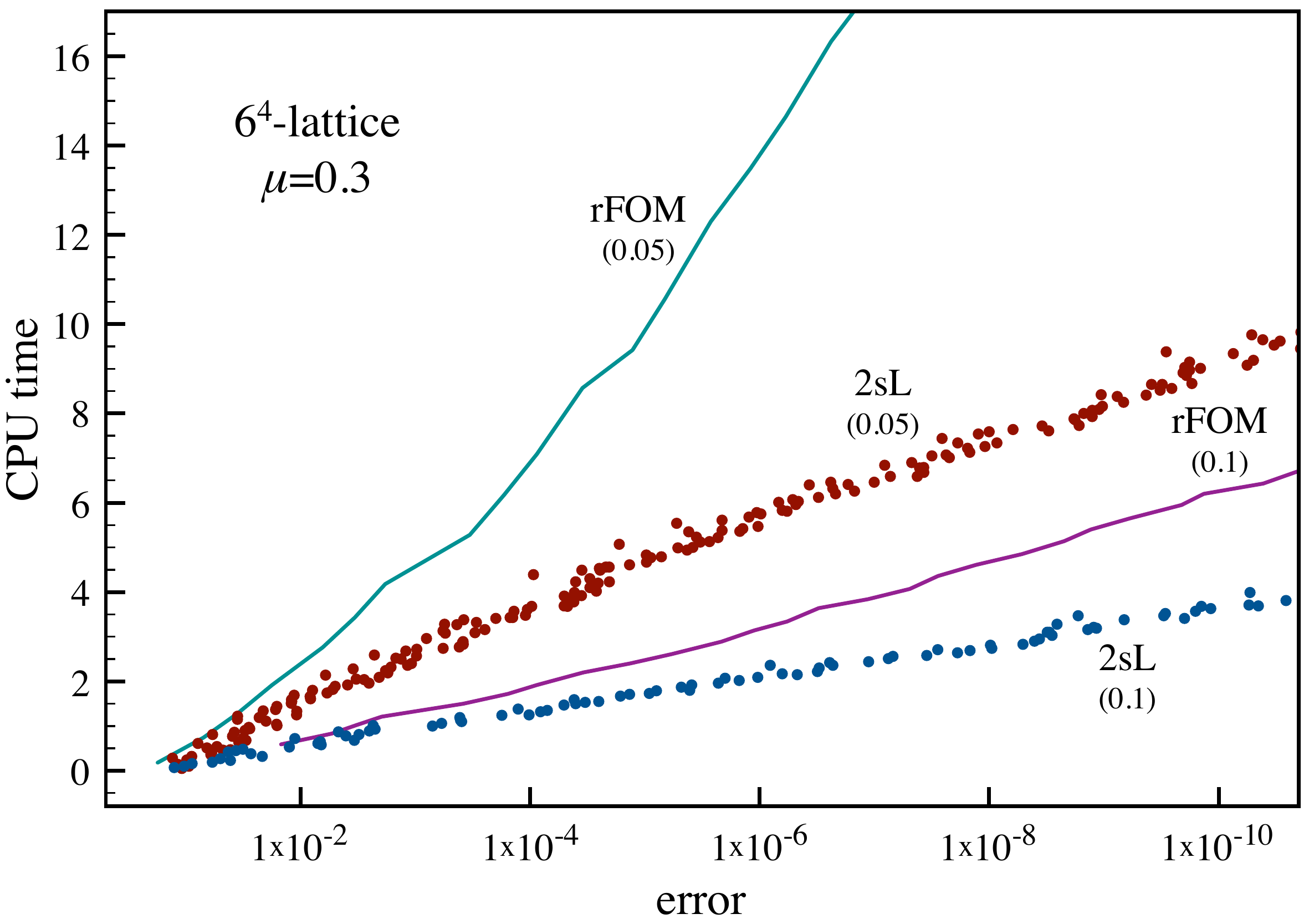}
\includegraphics[width=\figwidth,type=pdf,ext=.pdf,read=.pdf]{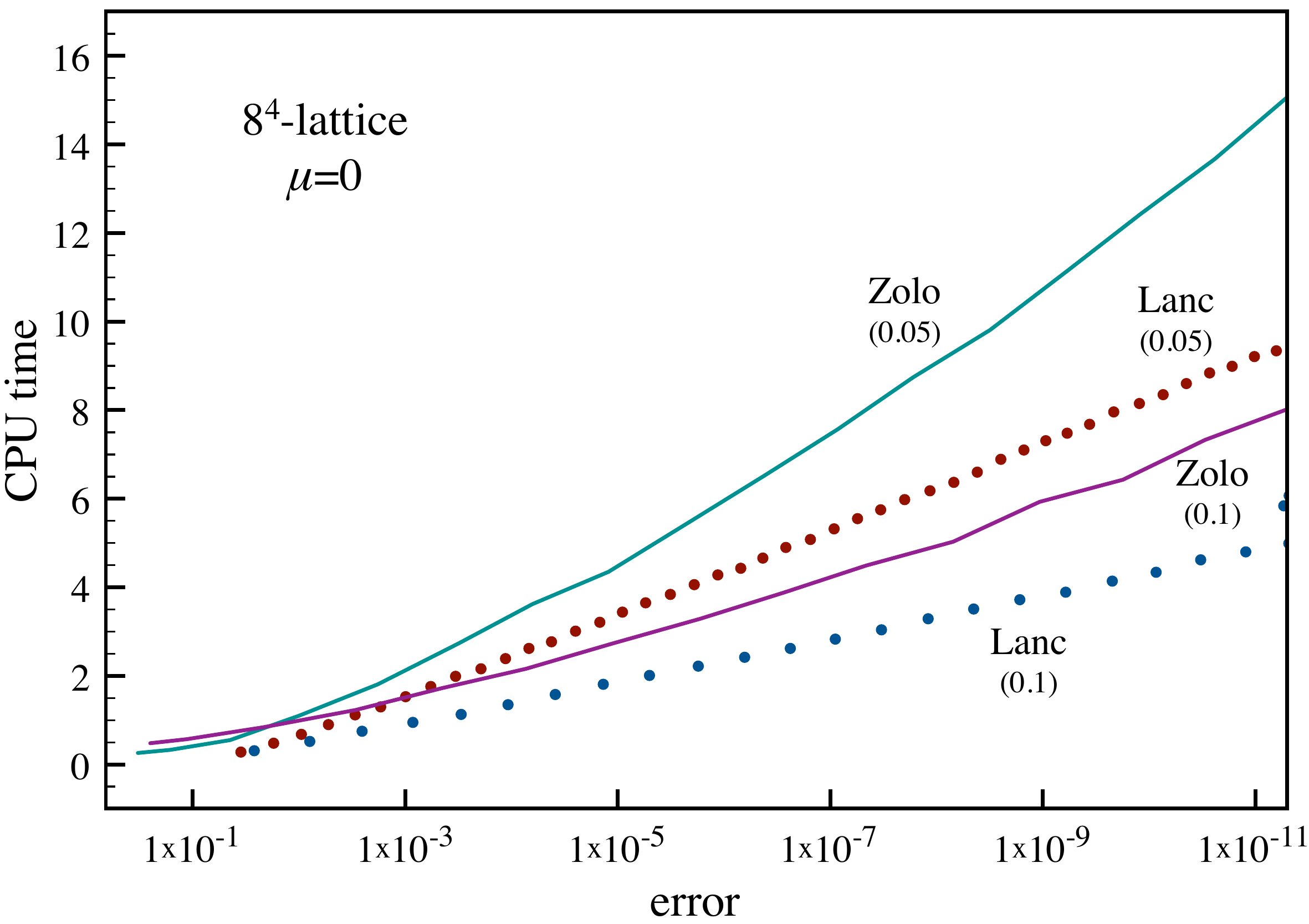}
\hspace{10mm}
\includegraphics[width=\figwidth,type=pdf,ext=.pdf,read=.pdf]{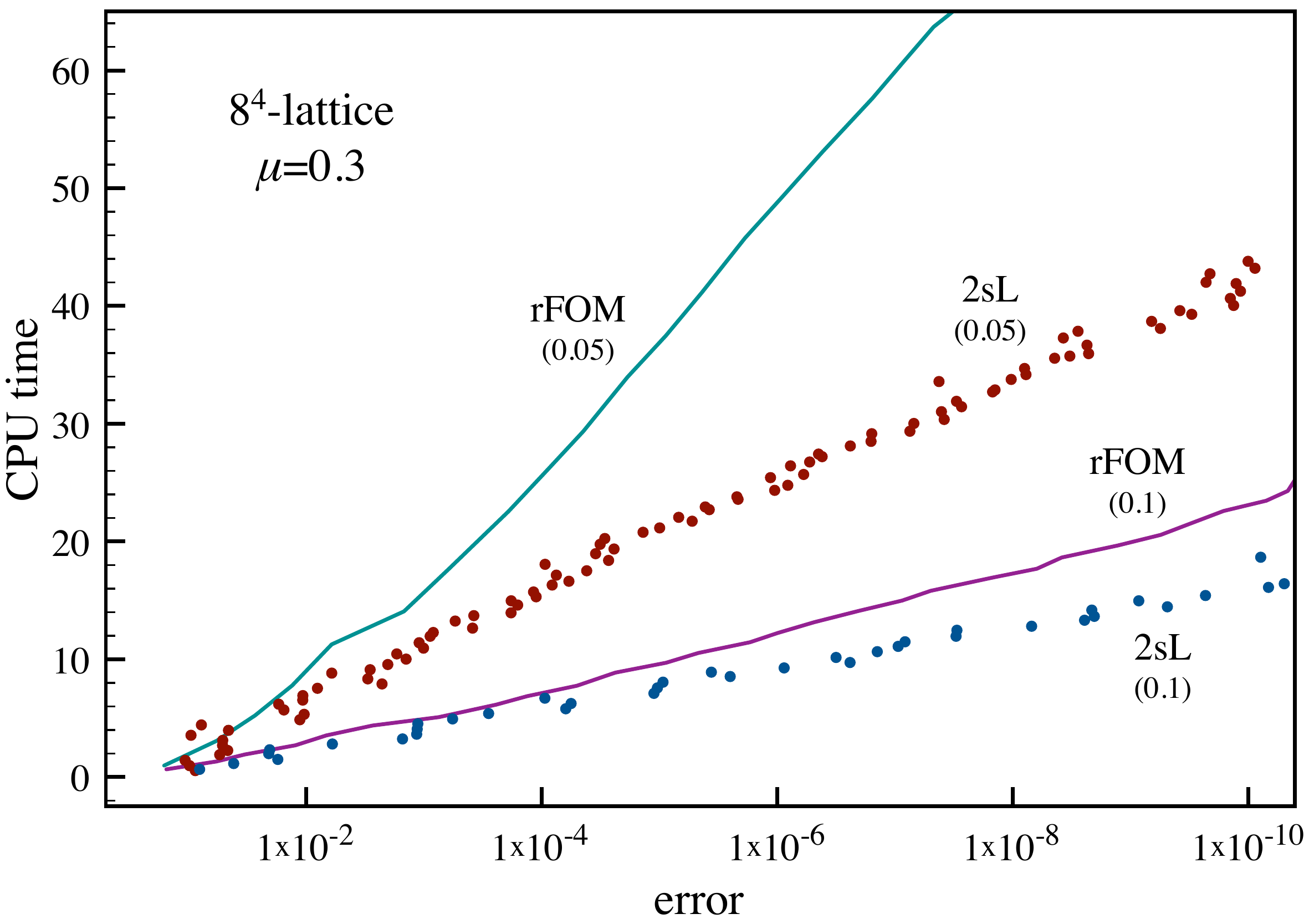}
\includegraphics[width=\figwidth,type=pdf,ext=.pdf,read=.pdf]{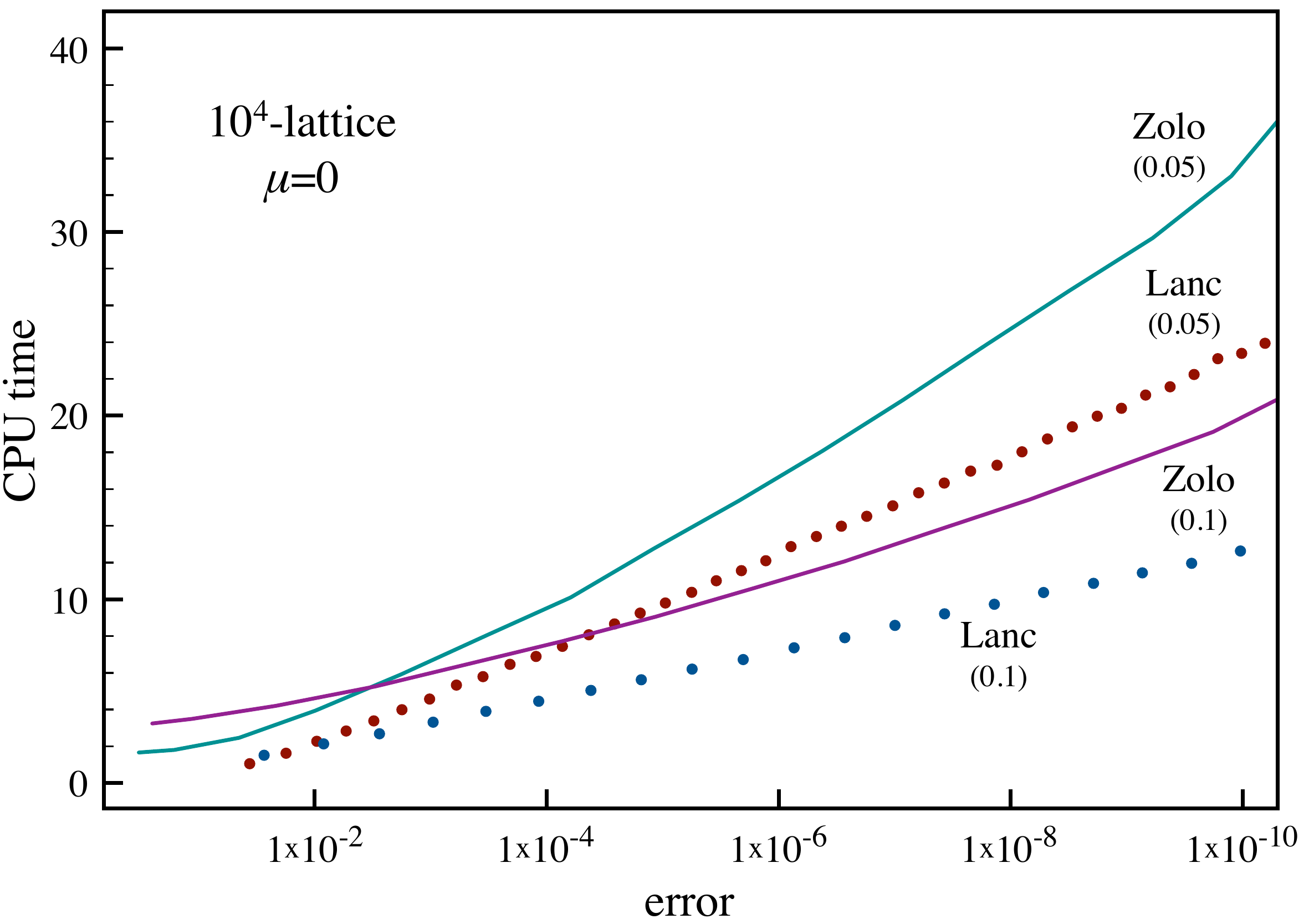}
\hspace{10mm}
\includegraphics[width=\figwidth,type=pdf,ext=.pdf,read=.pdf]{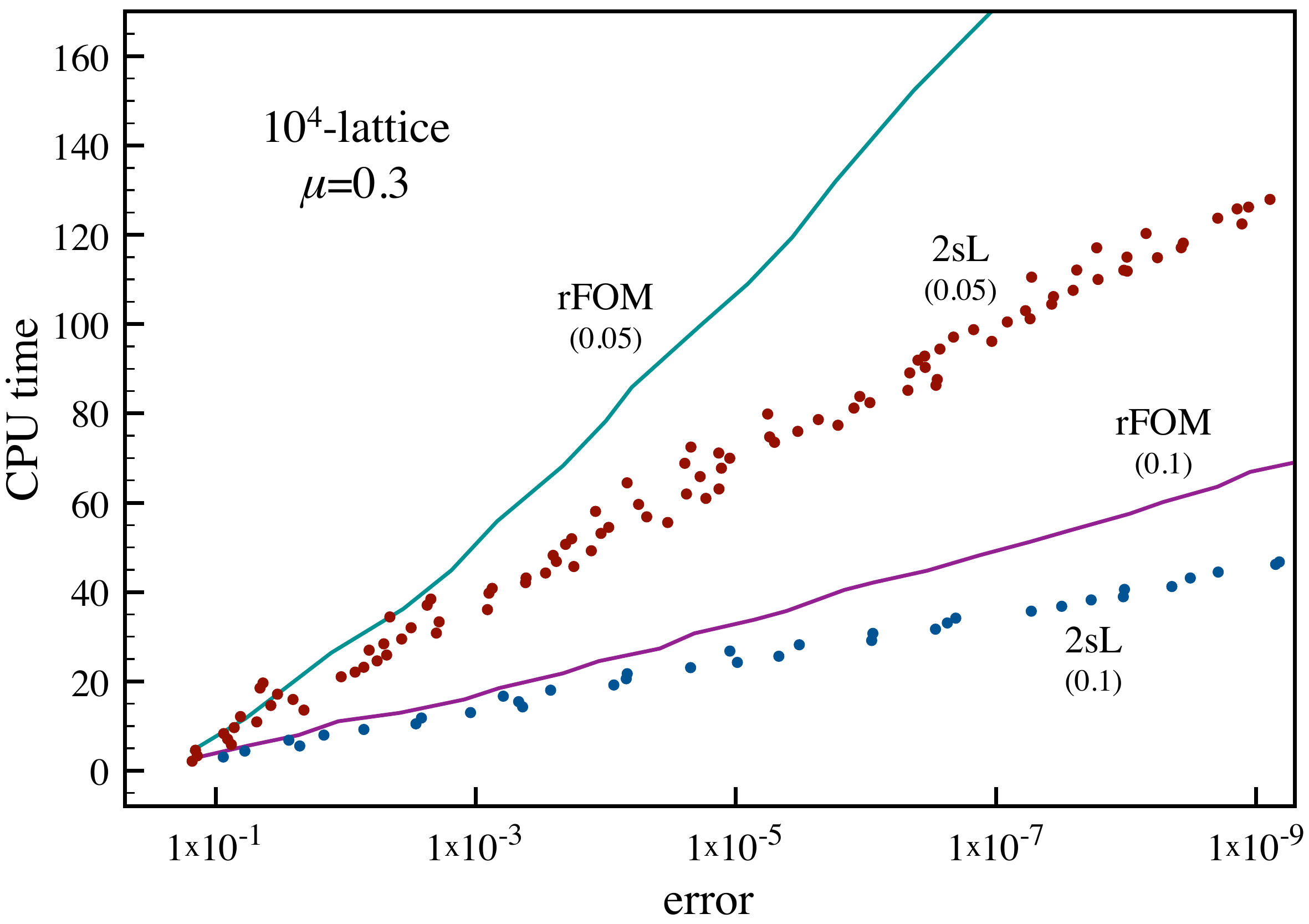}
\caption{Comparison of the nested Krylov subspace method (filled circles) with rational approximation methods (full lines) for lattices of sizes $4^4$, $6^4$, $8^4$ and $10^4$ for two different deflation gaps (given in parenthesis). Left: Hermitian case comparing the nested Lanczos approximation (Lanc) with the Zolotarev approximation (Zolo), evaluated using the Chroma QCD library. Right: non-Hermitian case with $\mu=0.3$ comparing the nested two-sided Lanczos method (2sL) with the Neuberger approximation evaluated with a restarted FOM algorithm (rFOM). 
The timings were measured on a single 2.4 GHz Intel Core 2 core with 8 GB of memory.}
\label{fig:timings}
\end{figure}

\begin{figure}[t]
\includegraphics[width=.49\textwidth]{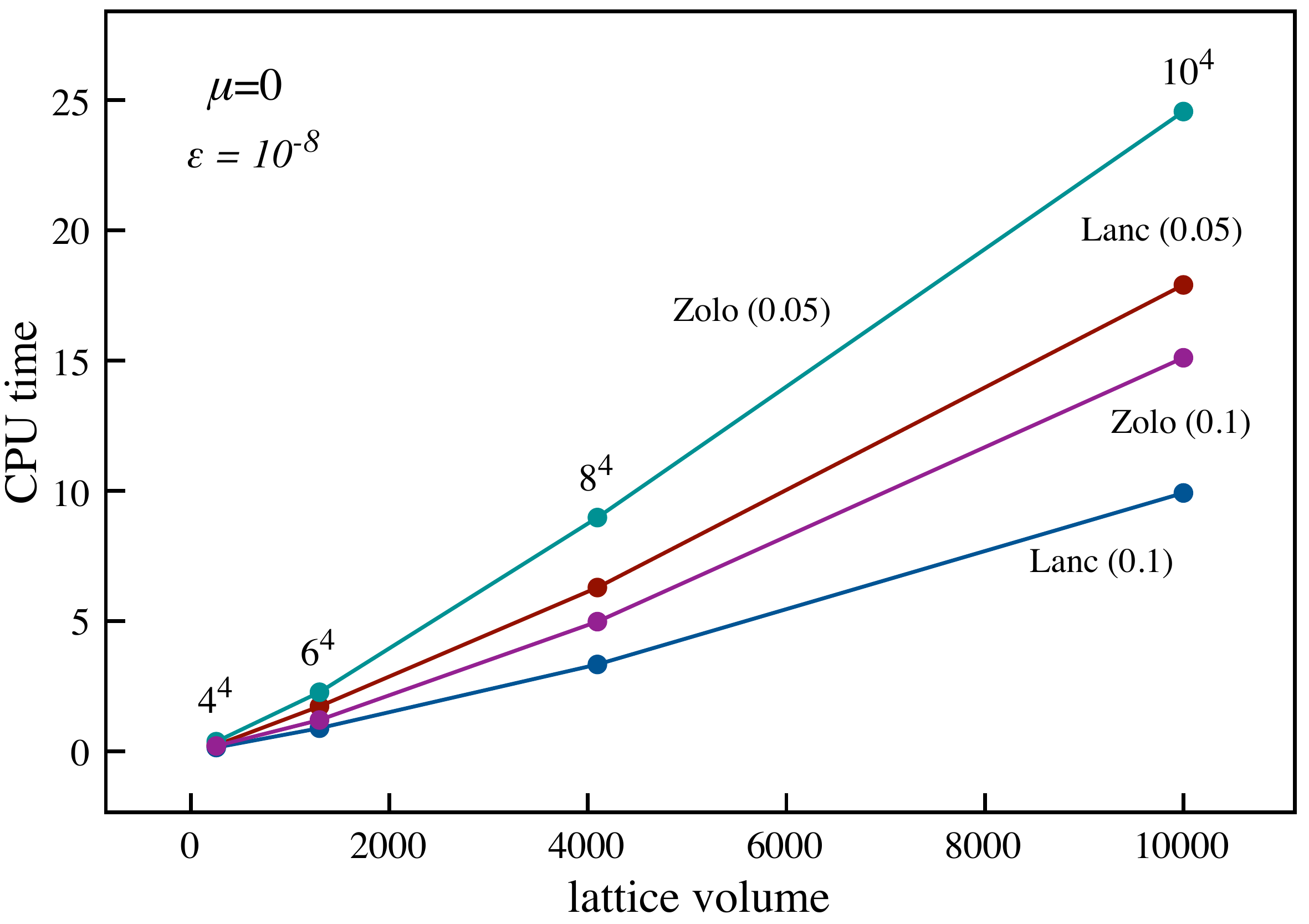}
\hfill
\includegraphics[width=.49\textwidth]{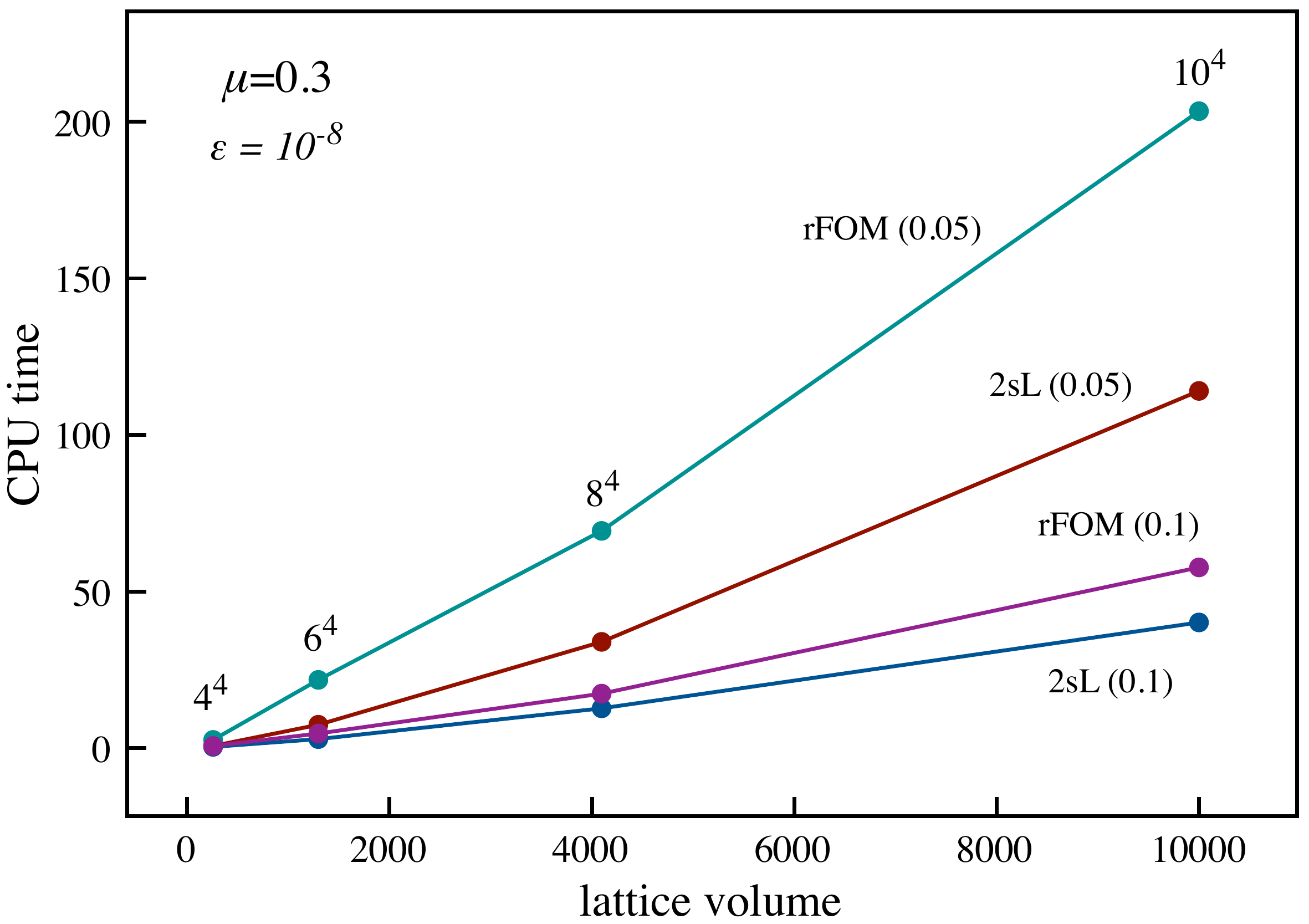}
\caption{Volume dependence of the run times for the nested method and rational approximation methods for the Hermitian (left) and non-Hermitian (right) case. The data are taken from Fig.~\ref{fig:timings} at an accuracy of $\varepsilon=10^{-8}$. }
\label{fig:time_vs_zolo}
\end{figure}

To evaluate the nested method further, we compare it to state-of-the-art rational approximation methods. In the Hermitian case the Zolotarev rational approximation, evaluated with a multi-shift conjugate gradient inverter \cite{vandenEshof:2002ms}, is routinely used in lattice simulations. In the non-Hermitian case, i.e., simulations at nonzero baryon density, overlap fermions are not yet commonly used because of their high cost, but recently an efficient algorithm was presented, which evaluates the Neuberger rational approximation using a multi-shift restarted FOM inverter \cite{Bloch:2009in}. 
In Fig.~\ref{fig:timings} we compare the results obtained with the nested Krylov subspace and rational approximation methods, and show how the CPU time varies as a function of the achieved accuracy for various lattice sizes. 
In all cases the Hermitian and non-Hermitian versions of the nested method perform better than the rational approximation method. 
The volume dependence of the run time for a fixed accuracy $\varepsilon$ can be extracted from Fig.~\ref{fig:timings} and is displayed for $\varepsilon=10^{-8}$ in Fig.~\ref{fig:time_vs_zolo}. Fits to the nested method results show a volume dependence which is slightly steeper than linear, i.e., proportional to $V^{1.2}$ for the Hermitian case and $V^{1.3}$ for the non-Hermitian case. The comparisons clearly demonstrate the good efficiency of the nested method. 

\subsection{Note on the memory usage}
\label{Sec:Memory}

In the numerical tests we observed that, for a fixed deflation gap, the Krylov subspace size needed to achieve a certain accuracy is almost independent of the lattice volume in the Lanczos approximation and only grows slowly with the volume in the two-sided Lanczos approximation. Therefore, the memory consumed by the Krylov basis $V_{\kout}$ is roughly proportional to the lattice volume.
For large lattice sizes this storage requirement might become too large to run the Krylov-Ritz approximation on a single node.

One solution, which only requires little storage, is to implement a \textit{double-pass} version of the algorithm, which is possible due to the use of short recurrences. 
In double-pass mode only the two most recently generated basis vectors are stored  
during the construction of the outer Krylov subspace basis. In the first pass the matrix $H_{\kout}$ is built and the product $\sgn(H_{\kout})e_1^{(\kout)}$ is computed with Eq.~\eqref{innersign}. In the second pass the basis vectors of the outer Krylov subspace are generated again and immediately added in a linear combination, whose coefficients were computed in the first pass.
The drawback of this variant is that the Krylov basis is constructed twice, such that the corresponding CPU time will be doubled. 

The more efficient solution is to parallelize the single-pass version of the algorithm, such that the memory requirement gets distributed over the available nodes. Benchmarks on larger volumes, using such a parallel implementation, are currently being performed.

\subsection{Multi-level nesting}

\begin{figure}
\centering
\includegraphics[width=0.49\textwidth]{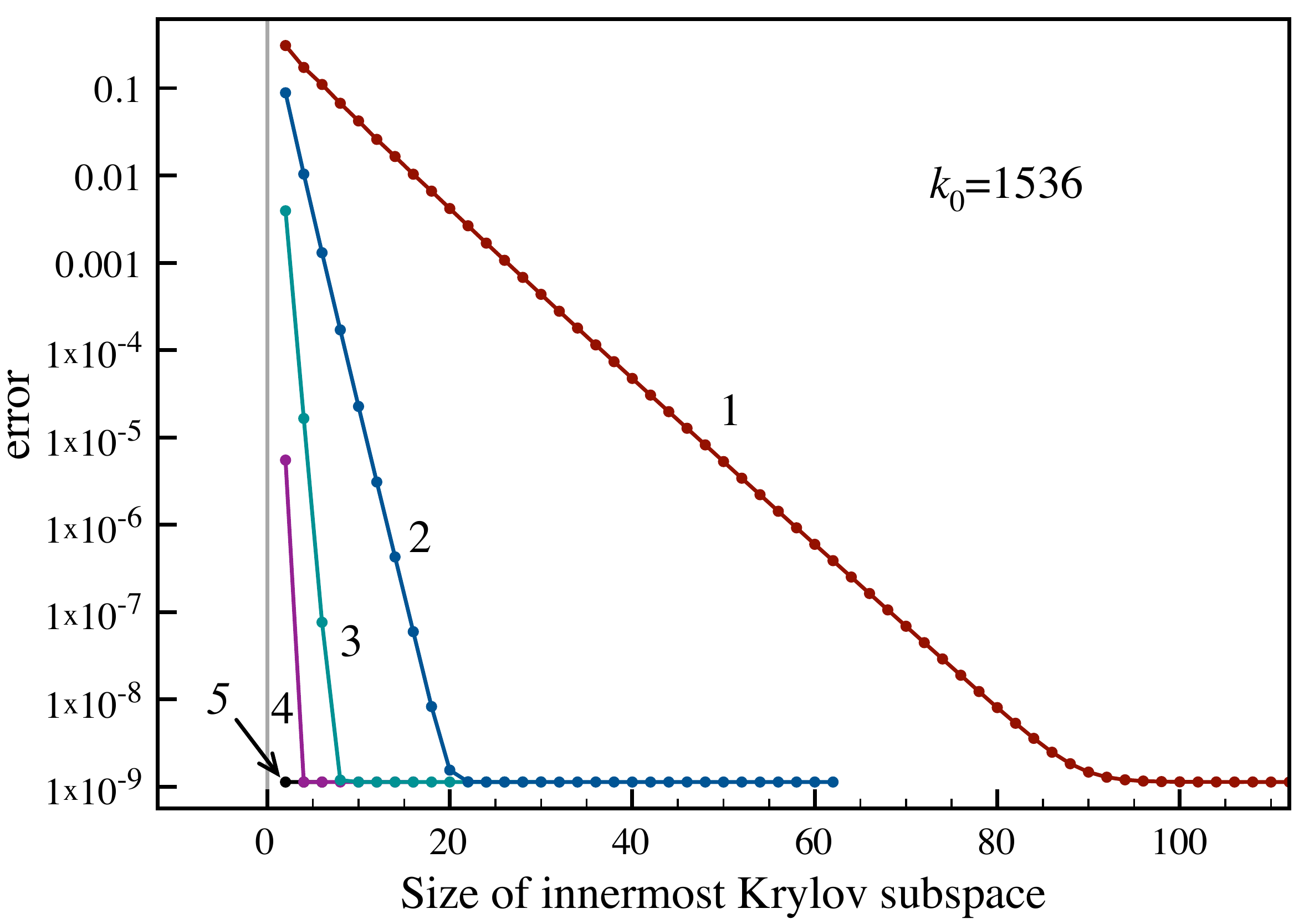}
\caption{Accuracy $\varepsilon$ of the nested method with 1, 2, 3, 4 and 5 nesting levels for lattice size $8^4$ in the Hermitian case with deflation gap $\Delta=0.055$ and $k_0=1536$. We plot the dependence of $\varepsilon$ on the size $k_i$, $i=1,\ldots,5$, of the innermost Krylov space. The convergence curves are labelled with the number of levels in the method. For the $i$-level method, the outer levels $k_j$, $j=1,\ldots,i-1$ are fixed to a value in their convergence region. }
\label{fig:8888_all_k_dependence}
\end{figure}

In principle, if the inner Krylov subspace in Eq.~\eqref{nested} is still too large for an efficient application of the RHi on the inner Ritz matrix, the nested method could be applied recursively.\footnote{Note that for all cases considered in the current study a single level of nesting was sufficient.} In this case we rename $\kout$ to $k_0$, $\kin$ to $k_1$, and add more recursively nested levels $k_i$ as necessary. Except for the deepest level, the matrix-vector product $\sgn(H_{k_i})e_1^{(k_i)}$ required at level $i$ will be computed with a Krylov-Ritz approximation \eqref{innersign} in the nested Krylov subspace $\mathcal{K}_{k_{i+1}}(H_{k_i}',e_1^{(k_i)})$, where $H_{k_i}'$ is defined by Eq.~\eqref{precond} on $H_{k_i}$ and typically $k_{i+1} \ll k_i$. At the deepest level the sign function of the Ritz matrix will be evaluated with the RHi. This multi-level nesting is illustrated in Fig.~\ref{fig:8888_all_k_dependence}, where we show the convergence curves for 1, 2, 3, 4 and 5 nested levels as a function of the size of the innermost Krylov subspace, with the sizes of all outer levels kept fixed to some value inside their convergence region (the convergence curves do not depend on the precise choice of the outer $k_i$'s).
As before, the convergence criterion is set by the size  $k_0$ of the outer Krylov subspace. 
 Each additional level lowers the size of the Krylov subspace. In the case depicted in Fig.~\ref{fig:8888_all_k_dependence} the optimal Krylov subspace sizes, i.e. where convergence is reached, for the successive levels decreases from $1536 \to 90 \to 20 \to 8 \to 4 \to 2$.
The improvement is most dramatic for the first nested level, but fast convergence is exhibited at all levels\footnote{For each level the $p$-factor for Eq.~\eqref{precond} is computed using Eq.~\eqref{popt}, using appropriately approximated boundaries for the spectrum of the Ritz matrix of the previous level. Note that the factor $p$ converges to $1$ as more levels are introduced, and the preconditioning step converges to the RHi.}. This can be related to the quadratically convergent RHi, as the preconditioning step at each level mimics a step of the RHi and compresses the spectrum more and more towards $\pm 1$. Moreover, the judicious choice of $p$ at each level improves the convergence even more.
It is intriguing to note that, in the example of Fig.~\ref{fig:8888_all_k_dependence}, the sign of a matrix of dimension $n=49152$ can be evaluated to an accuracy of $10^{-9}$ by computing the sign of a $2 \times 2$ matrix, which is then lifted back to the original $n$-dimensional space through linear combinations of Krylov vectors. This emphasizes again the power of Krylov subspace methods.
 
\section{Conclusions}
\label{Sec:conclusions}

In this paper we have presented a nested Krylov subspace method which boosts the Krylov-Ritz approximations used to compute the sign function of both Hermitian and non-Hermitian matrices. The Krylov-Ritz approximation projects the matrix on a Krylov subspace in which it computes the sign function exactly, before lifting it back to the original space.
Its standard implementation suffers from the CPU intensive computation of the sign of the Ritz matrix, which goes like the cube of the Krylov subspace size. 
By making an additional projection on a much smaller Krylov subspace, the nested method significantly reduces the total computation time of the Krylov-Ritz approximation, without affecting its accuracy.
Numerical tests showed that the nested method works equally well for Hermitian and non-Hermitian matrices and is more efficient than state-of-the-art rational approximation methods. Moreover, it exhibits a good, close to linear, volume scaling. We are currently investigating the efficiency of the nested method for larger lattice volumes using a parallel implementation of the algorithm.

To end, we comment on the relation between the nested method and the extended Krylov subspace methods introduced in Ref.~\cite{extended-krylov}. 
An extended Krylov space is defined as
\begin{align}
    \mathcal{K}_k(A,A^{-1},x) = \myspan(x, Ax, A^{-1}x, A^2 x, A^{-2}x, \dotsc, A^{k-1}x, A^{-k+1}x),
\end{align}
and an approximation in that subspace approximates $f(A)$ by the sum $Q(A) = \sum_{-k+1}^{k-1} c_i A^i$. 
In the nested method we construct the $\kin$-dimensional Krylov subspace $\mathcal{K}_{\kin}(H_{\kout}',e_1^{(\kout)})$, which forms an $\kin$-dimensional subspace of the $(2\kin-1)$-dimensional extended Krylov subspace $\mathcal{K}_{\kin}(p H_{\kout},(pH_{\kout})^{-1},e_1^{(\kout)})$.
The nested method implicitly fixes the coefficients of the positive and negative powers of $Q(H_{\kout})$ to be equal, $c_{-i} = c_i$,  which follows from the use of the property $\sgn (H_k) = \sgn(H_k+H_k^{-1})$.
Hence, the nested method implicitly truncates the size of the extended Krylov subspace. 

Approximations for the sign function in extended Krylov subspaces have been briefly considered recently \cite{Knizhnerman2010}, however not in combination with the nesting of Krylov subspaces, i.e. the extended subspace is constructed for the original matrix $A$, not for the Ritz matrix $H_k$. Evidently this is not feasible in the application to lattice QCD as the inversion of the $\gamma_5$-Wilson Dirac operator is too expensive in order to construct extended Krylov subspaces. 

To conclude, we briefly consider the application of the nested method to other matrix functions. The method presented in Sec.~\ref{Sec:nested} requires a transformation which leaves the matrix function invariant, similar to Eq.~\eqref{precond} for the sign function. If such a transformation is not known, the nested method could be adapted by using an extended Krylov subspace method at the inner level. This is also a topic of work in progress.

\section*{Acknowledgements} 

We would like to thank Andreas Frommer and Tilo Wettig for useful discussions.

\appendix

\section{Lanczos algorithm}
\label{alg:lanczos}
\begin{algorithmic}
    \STATE $v_1 \leftarrow \frac{x}{\lvert x \rvert}$
    \STATE $r \leftarrow Av_1$
    \FOR{$j=1$ to $k$}
        \STATE $H(j,j) \leftarrow v_j^\dagger r$
        \STATE $r \leftarrow r-H(j,j)v_j$
        \IF{$j=k$}
            \STATE stop
        \ENDIF
        \STATE $\beta \leftarrow \sqrt{r^\dagger r}$
        \STATE $H(j,j+1) \leftarrow \beta$
        \STATE $H(j+1,j) \leftarrow \beta$
        \STATE $v_{j+1} \leftarrow \frac{1}{\beta}r$
        \STATE $r \leftarrow Av_{j+1}$
        \STATE $r \leftarrow r - \beta v_j$
    \ENDFOR
\end{algorithmic}
All $H(i,j)$ not assigned above are zero. Consequently $H$ is tridiagonal and symmetric. The $v_j$ are the column vectors of the matrix $V_k$.

\section{Two-sided Lanczos algorithm}
\label{alg:bilanczos}
\begin{algorithmic}
    \STATE $v_1 \leftarrow \frac{x}{\lvert x \rvert}$
    \STATE $w_1 \leftarrow v_1$
    \STATE $r \leftarrow Av_1$
    \STATE $l \leftarrow A^\dagger w_1$
    \FOR{$j=1$ to $k$}
        \STATE $H(j,j) \leftarrow w_j^\dagger r$
        \STATE $r \leftarrow r-H(j,j)v_j$
        \STATE $l \leftarrow l-(H(j,j))^* w_j$
        \IF{$j=k$}
            \STATE stop
        \ENDIF
        \STATE $\delta \leftarrow r^\dagger l$
        \IF{$\delta = 0$}
            \STATE serious breakdown, stop
        \ENDIF
        \STATE $\beta \leftarrow \sqrt{\delta}$
        \STATE $H(j+1,j) \leftarrow \beta$
        \STATE $\gamma \leftarrow \frac{\delta^*}{\beta}$
        \STATE $H(j,j+1) \leftarrow \gamma$
        \STATE $v_{j+1} \leftarrow \frac{1}{\beta}r$
        \STATE $w_{j+1} \leftarrow \frac{1}{\gamma^*}l$
        \STATE $r \leftarrow Av_{j+1}$
        \STATE $l \leftarrow A^\dagger w_{j+1}$
        \STATE $r \leftarrow r - \gamma v_j$
        \STATE $l \leftarrow l - \beta^* w_j$
    \ENDFOR
\end{algorithmic}
The $v_j$ and $w_j$ are the column vectors of the matrices $V_k$ and $W_k$, respectively. All $H(i,j)$ not assigned above are zero. Consequently $H$ is tridiagonal, but not symmetric as in the Hermitian case. The coefficients $\beta$ and $\gamma$ are, non-uniquely, chosen to satisfy the biorthonormality condition
\begin{align}
    w_j^\dagger v_i = \delta_{ij}.
\end{align}
There are potential problems in the two-sided Lanczos process, namely serious breakdowns and near breakdowns, where $\delta \leftarrow r^\dagger l = 0$, respectively $\approx 0$, however, these were not encountered in our numerical tests.

\section{Nested algorithm}
\label{alg:nested}
Given a (non-)Hermitian matrix $A$, a source vector $x$ and the critical eigenvectors $r_i$ (left and right eigenvectors $l_i$ and $r_i$), with eigenvalues $\lambda_i$, $i=1,\dotsc,m$, do:
\begin{enumerate}
    \item Apply Left-Right deflation (see Ref.\cite{Bloch:2007aw}) to construct $x_{\ominus}$, where the components of the source vector $x$ along the eigenvectors $r_i$ have been removed:
        \begin{align*}
            x_{\ominus} &= x - \sum_{i=1}^m \langle l_i,x \rangle r_i,
        \end{align*}
        where $l_i=r_i$ for Hermitian $A$.
    \item Run the (two-sided) Lanczos algorithm from \ref{alg:lanczos} (\ref{alg:bilanczos}) with $A$ and $x_{\ominus}$ to obtain $V_{\kout}$ and $H_{\kout}$.
    
\item Perform an $LU$ decomposition of $p H_{\kout}$, e.g., with the LAPACK routine dgttrf (zgttrf). This yields a lower triangular matrix $L$ with unit diagonal and one sub-diagonal, and an upper triangular matrix $U$ with one diagonal and two super-diagonals. All other entries of $L$ and $U$ are zero.  

    \item Run the (two-sided) Lanczos algorithm with $H_{\kout}'=(p H_{\kout}+(p H_{\kout})^{-1})/2$ and source vector $e_1^{(\kout)}$ to construct the Krylov basis $V_{\kin}$ and the Ritz matrix $H_{\kin}$. 
     To do so apply $H_{\kout}'$ to each Krylov vector $v$:
\begin{enumerate}
\item[(a)] Compute $(p H_{\kout})^{-1} v$ using a sparse LU back substitution, e.g., with the LAPACK routine dgttrs (zgttrs).

\item[(b)] Compute $(p H_{\kout}) v$ and add to the result of (a). This tridiagonal multiply and add can be done efficiently using the BLAS band-matrix-vector multiplication routine dsbmv (zgbmv).

\end{enumerate}

    \item Run the RHi (or any other suitable method to compute the sign function) on $H_{\kin}$ to obtain $\sgn(H_{\kin})$.
    \item The final approximation is then given by
        \begin{align*}
            \sgn(A)x &\approx \sum_{i=1}^m\sgn(\lambda_i)\langle l_i, x \rangle r_i +
            |x_\ominus| V_{\kout}V_{\kin}\sgn(H_{\kin})e_1^{(\kin)} .
        \end{align*}
\end{enumerate}
Note that steps (3-5) are done in real arithmetic in the Hermitian case.

\bibliographystyle{JHEP}
\bibliography{biblio}

\end{document}